# The Small Satellites of Pluto as Observed by *New Horizons*

H. A. Weaver,[1*] M. W. Buie,[2] B. J. Buratti,[3] W. M. Grundy,[4] T. R. Lauer,[5] C. B. Olkin,[2] A. H. Parker,[2] S. B. Porter,[2] M. R. Showalter,[6] J. R. Spencer,[2] S. A. Stern,[2] A. J. Verbiscer,[7] W. B. McKinnon,[8] J. M. Moore,[9] S. J. Robbins,[2] P. Schenk,[10] K. N. Singer,[2] O. S. Barnouin,[1] A. F. Cheng,[1] C. M. Ernst,[1] C. M. Lisse,[1] D. E. Jennings,[11] A. W. Lunsford,[11] D. C. Reuter,[11] D. P. Hamilton,[12] D. E. Kaufmann,[2] K. Ennico,[9] L. A. Young,[2] R. A. Beyer,[6,9] R. P. Binzel,[13] V. J. Bray,[14] A. L. Chaikin,[15] J. C. Cook,[2] D. P. Cruikshank,[9] C. M. Dalle Ore,[9] A. M. Earle,[14] G. R. Gladstone,[16] C. J. A. Howett,[2] I. R. Linscott,[17] F. Nimmo,[18] J. Wm. Parker,[2] S. Philippe,[19] S. Protopapa,[12] H. J. Reitsema,[2] B. Schmitt,[19] T. Stryk,[20] M. E. Summers,[21] C. C. C. Tsang,[2] H. H. B. Throop,[22] O. L. White,[9] A. M. Zangari[2]

[1]The Johns Hopkins University Applied Physics Laboratory, Laurel, MD 20723, USA.

[2]Southwest Research Institute, Boulder, CO 80302, USA.

[3]National Aeronautics and Space Administration (NASA) Jet Propulsion Laboratory, California Institute of Technology, Pasadena, CA 91109, USA.

[4]Lowell Observatory, Flagstaff, AZ 86001, USA.

[5]National Optical Astronomy Observatory, Tucson, AZ 26732 USA.

[6]Search for Extraterrestrial Intelligence Institute, Mountain View, CA 94043, USA.

[7]Department of Astronomy, University of Virginia, Charlottesville, VA 22904, USA.

[8]Department of Earth and Planetary Sciences, Washington University, St. Louis, MO 63130, USA.

[9]NASA Ames Research Center, Space Science Division, Moffett Field, CA 94035, USA.

[10]Lunar and Planetary Institute, Houston, TX 77058, USA.

[11]NASA Goddard Space Flight Center, Greenbelt, MD 20771, USA.

[12]Department of Astronomy, University of Maryland, College Park, MD 20742, USA.

[13]Massachusetts Institute of Technology, Cambridge, MA 02139, USA.

[14]University of Arizona, Tucson, AZ 85721, USA.

[15]Independent Science Writer, Arlington, VT USA.

[16]Southwest Research Institute, San Antonio, TX 78238, USA.

[17]Stanford University, Stanford, CA 94305, USA.

[18]University of California, Santa Cruz, CA 95064, USA.
[19]Université Grenoble Alpes, CNRS, IPAG, F-38000 Grenoble, France.
[20]Roane State Community College, Oak Ridge, TN 37830, USA.
[21]George Mason University, Fairfax, VA 22030, USA.
[22]Planetary Science Institute, Tucson, AZ 85719, USA.

*To whom correspondence should be addressed. E-mail: hal.weaver@jhuapl.edu

**Abstract:** The *New Horizons* mission has provided resolved measurements of Pluto's moons Styx, Nix, Kerberos, and Hydra. All four are small, with equivalent spherical diameters of ~40 km for Nix and Hydra and ~10 km for Styx and Kerberos. They are also highly elongated, with maximum to minimum axis ratios of ~2. All four moons have high albedos (~50-90%) suggestive of a water-ice surface composition. Crater densities on Nix and Hydra imply surface ages $\gtrsim$ 4 Ga. The small moons rotate much faster than synchronous, with rotational poles clustered nearly orthogonal to the common pole directions of Pluto and Charon. These results reinforce the hypothesis that the small moons formed in the aftermath of a collision that produced the Pluto-Charon binary.

**Background and Context**

All of Pluto's four small moons Styx, Nix, Kerberos, and Hydra (in order of increasing distance from Pluto; hereafter we refer to this sequence as SNKH) were discovered using the *Hubble Space Telescope* (*HST*): Nix and Hydra in 2005 (*1*), Kerberos in 2011 (*2*), and Styx in 2012 (*3*). SNKH orbit the Pluto system barycenter in essentially the same plane (coincident with the Pluto-Charon orbital plane) and in nearly circular orbits with orbital semi-major axes of 42,656, 48,694, 57,783, and 64,738 km, and orbital periods of 20.2, 24.9, 32.2, and 38.2 days, respectively (*4,5*). These orbital periods are nearly integer multiples of Charon's 6.4 day orbital period, with ratios of 3:4:5:6 for SNKH, respectively (*4,5*). Sensitive searches for new moons with *New Horizons* were unsuccessful (*6*), demonstrating that no other moons larger than ~1.7 km in diameter (assuming the geometric albedo is ~0.5) are present at orbital radii between 5000 and 80,000 km, with less stringent limits at larger radii.

The long-term dynamical stability of Kerberos places severe constraints on the allowable masses and surface reflectances (i.e., albedos) of Nix and Hydra (*7*). These latter constraints, together with the disk-integrated brightness measurements of Nix and Hydra (brightness is proportional to the product of the object's cross-sectional area and its albedo), suggested that Nix and Hydra are relatively small, icy satellites (*7*). Adopting the hypothesis that impact-generated debris from the small moons produced a regolith covering Charon's surface (*8,9*), so that all of Pluto's moons would have the same visible light albedo (the visible light albedo of Charon is ~0.38), the average spherical-equivalent spherical diameters of Styx, Nix, Kerberos, and Hydra would be approximately 7, 40, 10, and 45 km, respectively. However, the observed brightness of Kerberos, together with dynamical constraints on its mass (*4*), suggested (*5*) it has a much lower albedo and a much larger size (e.g., 25 km diameter for an albedo of 0.06). Using an extensive set of *HST* brightness measurements over time, it was argued that Nix is highly elongated, with a maximum to minimum axial ratio of ~2 (*5*). Similar measurements suggested Hydra was also elongated, but less so than Nix (*5*). No stable rotational period could be found for either Nix or Hydra, perhaps suggesting that both bodies were tumbling chaotically as a result of the large and regular torques exerted on them by the Pluto-Charon binary (*5*).

The *New Horizons* mission provided an opportunity to make spatially resolved observations of Pluto's small moons, thereby testing the findings from Earth-based observations (*4,5*) and various theoretical predictions (*7,8,9*), by giving direct measurements of their sizes, shapes, surface albedo and color variations, and snapshots of their rotational states. In addition, an extensive and systematic set of unresolved panchromatic brightness measurements of the small moons over several months (early-April to early July 2015) was obtained by *New Horizons* during the approach to Pluto, which provides further information on their shapes and more precise information on their rotational states. Here we report on the results from the *New Horizons* observations of Pluto's small moons using data received on Earth by mid-December 2015. All the data discussed here were obtained by either the LOng Range Reconnaissance Imager (LORRI), which is a panchromatic camera (*10*), or the Multi-spectral Visible Imaging Camera (MVIC), which is a color camera (*11*). Infrared spectral measurements from the Linear Etalon Imaging Spectral Array (LEISA; *11*) will provide detailed compositional information, but those data will not be sent to Earth until March or April 2016.

***Physical Properties***

Table 1 presents a log of all the resolved measurements of Pluto's small moons. Some examples of the resolved images are shown in Fig. 1 (see also S1 and S7-S14). Systematic measurements of the brightness variations of Pluto's moons between May and early July 2015 were used together with the resolved measurements to constrain the sizes, shapes, rotation periods, and rotation poles of all four moons (*12*; see Table 2). Fig. 2 shows the observed brightness variations after phasing by the best-fit rotational periods (see also S2, S3). Unlike the case for Pluto and Charon, which each synchronously rotate with a period of ~6.4 days which equals their mutual orbital period around their common barycenter, Pluto's small moons rotate surprisingly rapidly (the fastest is ~10 hr for Hydra) and all are far from synchronous. The rotational poles (Table 2, S4) are clustered nearly orthogonal to the direction of the common rotational poles of Pluto and Charon: the inclination angles relative to the Pluto-Charon pole direction are 91°, 123°, 96°, and 110° for SNKH, respectively. Nominally, all the small moons have retrograde rotation, but Nix is the only one significantly so (i.e., retrograde with greater than 1σ confidence). This collection of inclinations is inconsistent with an isotropic distribution − even with only four points, a Kolmogorov-Smirnov test shows there is a less than 1% probability this a uniform distribution in inclination (i.e., the pole inclinations are non-uniform to a 2-3σ confidence level). These results on the rotational properties have not been seen in other regular satellite systems in the Solar System. Rapid rotations and large obliquities imply that tidal de-spinning has not played a major role in the moons' rotational histories. The moons have probably never reached the state of near-synchronicity where chaotic perturbations by Charon have been predicted to dominate (*5*); determining whether chaos plays a role in the moons' current rotational dynamics is deferred to a future study.

Pluto's small moons have highly elongated shapes with maximum to minimum axial ratios of ~1.5−2 (Table 2). Highly asymmetrical shapes are typical of many other small bodies in the Solar System and presumably reflect a growth process by agglomeration of small objects into loosely bound, macro-porous bodies whose gravity was insufficient to pull them into more spherical shapes. Kerberos, in particular, has a double-lobed shape, suggesting the merger of two smaller bodies. Hydra also has a highly asymmetrical shape that may also indicate the merger of smaller bodies, but the divots in Hydra's surface may plausibly have been produced by impacts from the local Kuiper belt population. The non-spherical shapes of Pluto's small satellites are consistent with their formation in the remnant disk produced by the collision of two large Kuiper belt objects (KBOs) that formed the Pluto-Charon binary (*13,14,15*).

Large uncertainties in the masses of the small moons (up to ~100%), as well as significant uncertainties in their volumes, precludes determining accurate values for their densities at this time (densities of 0 are within the current error estimates). However, the *New Horizons* results on Kerberos (see below) clearly demonstrate that the current dynamical estimate for its mass (*4*) is an overestimate, possibly by a factor of ~100.

***Albedos and Surface Features***

The *New Horizons* spacecraft trajectory was designed to maximize the scientific return on Pluto and Charon, and observations of those bodies were given priority during the flyby. Nevertheless, Nix and Hydra were imaged with sufficient resolution to investigate brightness variations, color variations, and topographical features across their surfaces. Direct surface reflectance (I/F, the ratio of reflected intensity to incident flux from the Sun) measurements are also available for Kerberos and Styx. Contour maps of raw I/F values for the best resolved images of SNKH are shown in S8-S11.

All four small satellites have high albedos, similar to those of some of Saturn's small, icy moons (*16*). Even at a phase angle of 34°, I/F on Hydra reaches 0.56 ± 0.03 (the quoted error is the estimated ±1σ uncertainty in the LORRI absolute calibration). Peak values of I/F for SNK are 0.40 ± 0.02, 0.57 ± 0.03, and 0.45 ± 0.02 at phase angles of 17.3°, 9.5°, and 24.7°, respectively. Converting these reflectances into geometric albedos (which, by definition, is the I/F at a phase angle of 0°) requires knowledge of the phase function, including any brightness increase that might occur near 0° due to coherent backscattering. We used the total light integrated over the best image of Nix to constrain its geometric albedo, and then we used the relative albedos derived from the unresolved brightness measurements of all the small moons (when the phase angle was identical for all the small moons; we also accounted for rotational lightcurve variations among the small moons) to estimate the geometric albedos of SKH. We derive a visible band apparent magnitude ($V$) of $V = -0.79 \pm 0.05$ for the best image of Nix, which was taken at a moderate phase angle of 9.5°. For a phase law with a linear phase coefficient of 0.04 mag/deg, which we favor, the geometric albedo is 0.61. For a phase law of 0.02 mag/deg, which is near the extreme of what is observed for asteroids and other planetary satellites, the geometric albedo would be 0.51. Thus, for Nix we adopt a geometric albedo of 0.56 ± 0.05. Using the relative albedo measurements derived from the extensive set of observations taken during May-July 2015 (Fig. 2), we derive the geometric albedos for SKH listed in Table 2. None of these geometric albedo values account for potential rapid brightness increases near 0° phase angle, where observations from *New Horizons* were not possible.

The vast majority of small KBOs (but still usually larger than 100 km in diameter) have *V*-band geometric albedos < 20%, with typical values of ~10% (*17,18*). Thus, these new measurements provide further evidence that Pluto's small moons were not captured from the general Kuiper belt population, but instead formed by agglomeration in a disk of material produced in the aftermath of the Charon-forming collision (*13,14,15*). The geometric albedos of the small moons appear to be larger than the value for Charon (~0.38), contrary to the prediction that regolith transfer from the small moons to Charon would result in approximately equal geometric albedos for all of Pluto's moons (*8,9*). The high albedo and small size for Kerberos directly contradicts the prediction (*5*) that Kerberos should be large and dark. However, our observational results from *New Horizons* support the predictions from two theoretical studies (*7,15*), which argued for high albedos for the small moons based on dynamical considerations.

Although diagnostic compositional spectra on Pluto's small moons have not yet been received from the *New Horizons* spacecraft, the combination of high surface albedo and their residence at large heliocentric distances strongly suggest that all the moons are covered with icy material. By analogy with Charon, which is covered in $H_2O$ ice but not massive enough to retain more volatile ices (e.g., $N_2$, $CH_4$, CO, etc.) over the age of the Solar System (*19,20*), we propose it is likely that the surfaces of Pluto's small moons are also covered with $H_2O$ ice. We further note that if the Pluto-Charon binary was produced by a giant collision in which both precursor bodies were at least partially differentiated (possessing icy surface layers), any small moons formed in the resulting debris disk are predicted to be rich in water ice (*14,15*).

Both Nix and Hydra have surface features that we attribute to impact craters caused by bombardment from small bodies in the local Kuiper belt population. We identify 11 crater-like features on Nix and 3 crater-like features on Hydra (*12;* Figs. S12-S14). We calculate crater densities of $(1{-}3) \times 10^{-3}$ km$^{-2}$ for crater diameters ≥ 4 km, which match or exceed the values found (*21*) on the older regions of Pluto and Charon after accounting for the much lower gravities of Nix and Hydra (Fig. 3; see *22*). Given the resolution and phase angle limitations of the images (Table 1), these cumulative counts should be considered minimum values. Assuming our identification of craters on Nix and Hydra is correct, the high crater densities suggest (*21,22*) that the surfaces of Nix and Hydra date back at least ~4 Ga, when the density of the Kuiper belt was perhaps a hundred times larger than the present-day value (*23*). Catastrophic disruption since that time is not predicted for Nix and Hydra (*22*), which is consistent with their large surface ages.

Nix and Hydra have mostly neutral (i.e., grey) colors, but Hydra is somewhat bluer than Nix (Fig. 4). Perhaps Hydra's surface is icier than Nix's, which might explain both its higher albedo and bluer color. The largest crater on Nix's surface is redder than the rest of its surface (Fig. 5). Possible explanations include the impacting body having had a different composition to Nix, or the impact exposed material with a different composition from below Nix's surface. No significant color variation is associated with the other impact craters identified on Nix's or Hydra's surfaces.

***Implications***

The *New Horizons* observations of Pluto's small satellites have produced a number of results: rapid rotation rates and unusual pole orientations; bright, icy surfaces with albedos and colors distinctly different from those of Pluto and Charon; evidence of merged bodies; and surface ages of $\gtrsim$ 4 Ga. Perhaps the rotational properties of the small moons are affected by stochastic collisional processes, which could both spin up the moons and re-orient their rotational axes, more strongly than previously appreciated. The presence of a distinctly different layer of material uncovered and/or deposited by the impact on Nix demonstrates that these bodies possess regoliths. The *New Horizons* measurements suggest that regolith sharing between the small moons and Charon is less extensive than previously thought (8,9). Collisions are also implicated in determining the shapes of the small moons: Kerberos appears to record the slow merger of two separate bodies, and Hydra has large surface indentations that might reflect mass loss by impacting bodies. However, the major collisional evolution of the small moons was probably limited to the first several hundred million years after the Solar System's formation because the surface crater retention ages of Nix and Hydra are $\gtrsim$ 4 Ga.

**Acknowledgements**

We thank the many dedicated engineers who contributed to the success of the *New Horizons* mission and NASA's Deep Space Network for a decade of excellent support to *New Horizons*. This work was supported by NASA's *New Horizons* project. As contractually agreed to with NASA, fully calibrated *New Horizons* Pluto system data will be released via the NASA Planetary Data System at https://pds.nasa.gov/ in a series of stages in 2016 and 2017 due to the time required to fully downlink and calibrate the dataset. Dr. Hamilton acknowledges financial support from NASA's Origins research program, and Drs. Philippe and Schmitt acknowledge financial support from France's Centre National d'Etudes Spatiales. Alan Stern is also affiliated with Florida Space Institute, Uwingu LLC, Golden Spike Co., and World View Enterprises.

**Supplementary Materials**

www.sciencemag.org
Materials and Methods
Supplementary Text
Tables S1 – S4
Figs. S1 – S14
References (25 – 30)
Supplementary movie

**Table1. Log of available resolved observations of Pluto's small moons.** All observations of Pluto's small satellites with a resolution better than 15 km/pixel and downlinked to the Earth before 2015-December-15 are listed below. The dates are the mid-observation times at the *New Horizons* spacecraft, and the values in parentheses are the number of images containing the object in that Observation Name. The resolution refers to the projected distance at the object subtended by a single instrument pixel. The phase angle is the Sun-object-New Horizons angle. All observations were taken with the LORRI panchromatic camera (*10*), except "N_COLOR_2" and "N_MPAN_CA", which were taken with the MVIC color camera (*11*).

| Object | Observation Name | Date in 2015 (MM-DD HH:MM:SS UTC) (number of images) | Resolution (km/pixel) | Phase Angle (deg) |
|---|---|---|---|---|
| Styx | U_TBD_1_02 | 07-13 23:44:56 (6) | 3.13 | 17.3 |
| Nix | N_LORRI_APPR_1D2 | 07-13 23:19:29 (2) | 2.92 | 13.4 |
| | N_COLOR_2 | 07-14 08:03:29 (1) | 3.10 | 8.45 |
| | N_LORRI_BACKUP | 07-14 08:06:58 (4) | 0.76 | 8.34 |
| | N_COLOR_BEST | 07-14 09:12:59 (1) | 1.99 | 6.13 |
| | N_LEISA_LORRI_BEST | 07-14 10:03:10 (1) | 0.30 | 9.45 |
| | N_MPAN_CA | 07-14 11:16:35 (1) | 0.45 | 85.9 |
| | N_DEP_SOONEST | 07-14 14:56:57 (16) | 0.93 | 158 |
| Kerberos | U_TBD_2 | 07-14 04:24:17 (4) | 1.97 | 24.7 |
| Hydra | H_LORRI_APPR_1D2 | 07-13 23:16:23 (2) | 3.18 | 21.5 |
| | H_COLOR_1 | 07-14 04:54:09 (1) | 7.17 | 26.7 |
| | H_COLOR_BEST | 07-14 07:37:00 (1) | 4.59 | 33.5 |
| | H_LORRI_BEST | 07-14 07:40:28 (8) | 1.14 | 33.9 |

**Table 2. Properties of Pluto's Small Satellites.** The sizes (diameters) are 3-dimensional ellipsoidal shape best-fits to the resolved and unresolved (lightcurve) measurements (*12*). Uncertainties are ±3 km (±1σ) for Styx, Kerberos, and Nix and ±10 km (±1σ) for Hydra. Kerberos has a dual-lobed shape that is not fit well by a single ellipsoid. The orbital periods are from (*5*). The rotation rates are determined from analyses of lightcurve data taken over several months (*12*). Rotational pole directions are determined from a model that attempts to match both the lightcurve measurements and the resolved measurements (*12*). The pole positions listed below are accurate to ±10° (±1σ, also see S4); the rotational poles of Pluto and Charon both point at [RA,DEC] = [132.993°, −6.163°]. The geometric albedos listed here may not fully account for any potential rapid increase in brightness near 0° phase angle (see the text for further details). Based on a recent (November 2015) analysis of stellar calibration data, we have reduced LORRI's sensitivity by 20% compared to the pre-flight value, which raises the derived geometric albedo values (tabulated below) by 20% relative to the values based on the original calibration. LORRI's sensitivity has been stable at the ~1% level since launch, and a more definitive absolute calibration is expected from stellar observations planned in July 2016.

| Object | Size (km) | Orbital Period (days) | Rotation Rate (days) | Rotation Pole [RA,DEC] (deg) | Geometric Albedo |
|---|---|---|---|---|---|
| Styx | 16 x 9 x 8 | 20.16155 ± 0.00027 | 3.24 ± 0.07 | [196,61] | 0.65 ± 0.07 |
| Nix | 50 x 35 x 33 | 24.85463 ± 0.00003 | 1.829 ± 0.009 | [350, 42] | 0.56 ± 0.05 |
| Kerberos | 19 x 10 x 9 | 32.16756 ± 0.00014 | 5.31 ± 0.10 | [222,72] | 0.56 ± 0.05 |
| Hydra | 65 x 45 x 25 | 38.20177 ± 0.00003 | 0.4295 ± 0. 0008 | [257,-24] | 0.83 ± 0.08 |

**Figures and Captions**

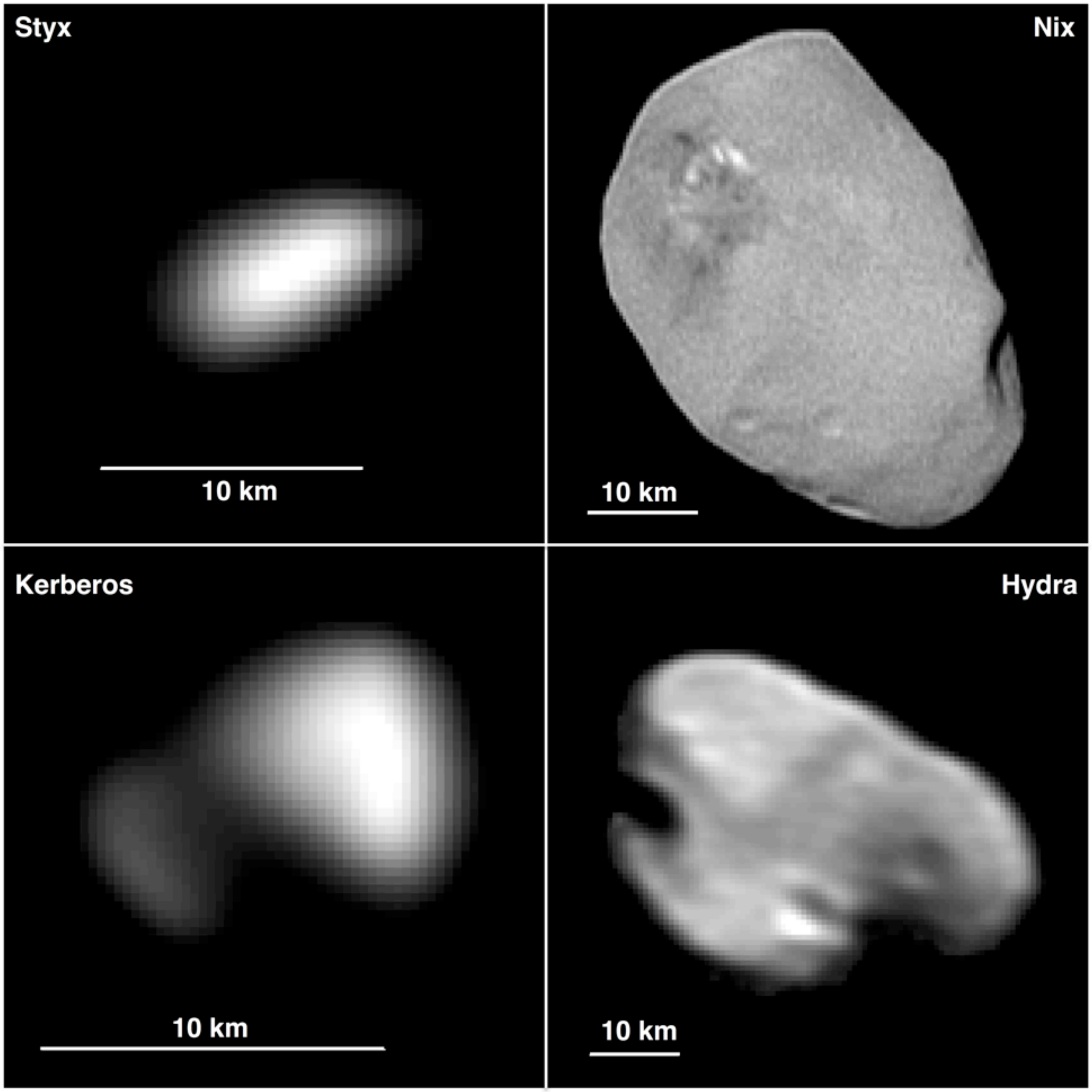

**Fig. 1. Best resolved images of Pluto's four small moons.** The images are displayed with celestial North up and East to the left. The Styx image is a deconvolved (*12*) composite of 6 images from U_TBD_1_02 (Table 1) that has been re-sampled with pixels 8 times smaller than the native pixel scale for cosmetic purposes. The Nix image is a deconvolved single image from N_LEISA_LORRI_BEST and is displayed with the native pixels. The Kerberos image is a deconvolved composite of 4 images from U_TBD_2 and has been re-sampled with pixels 8 times smaller than the native pixel scale for cosmetic purposes (*12*; S1). The Hydra image is a deconvolved composite of 2 images from H_LORRI_BEST with pixels 2 times smaller than the native scale. Some surface features on Nix and Hydra appear to be impact craters (*12*).

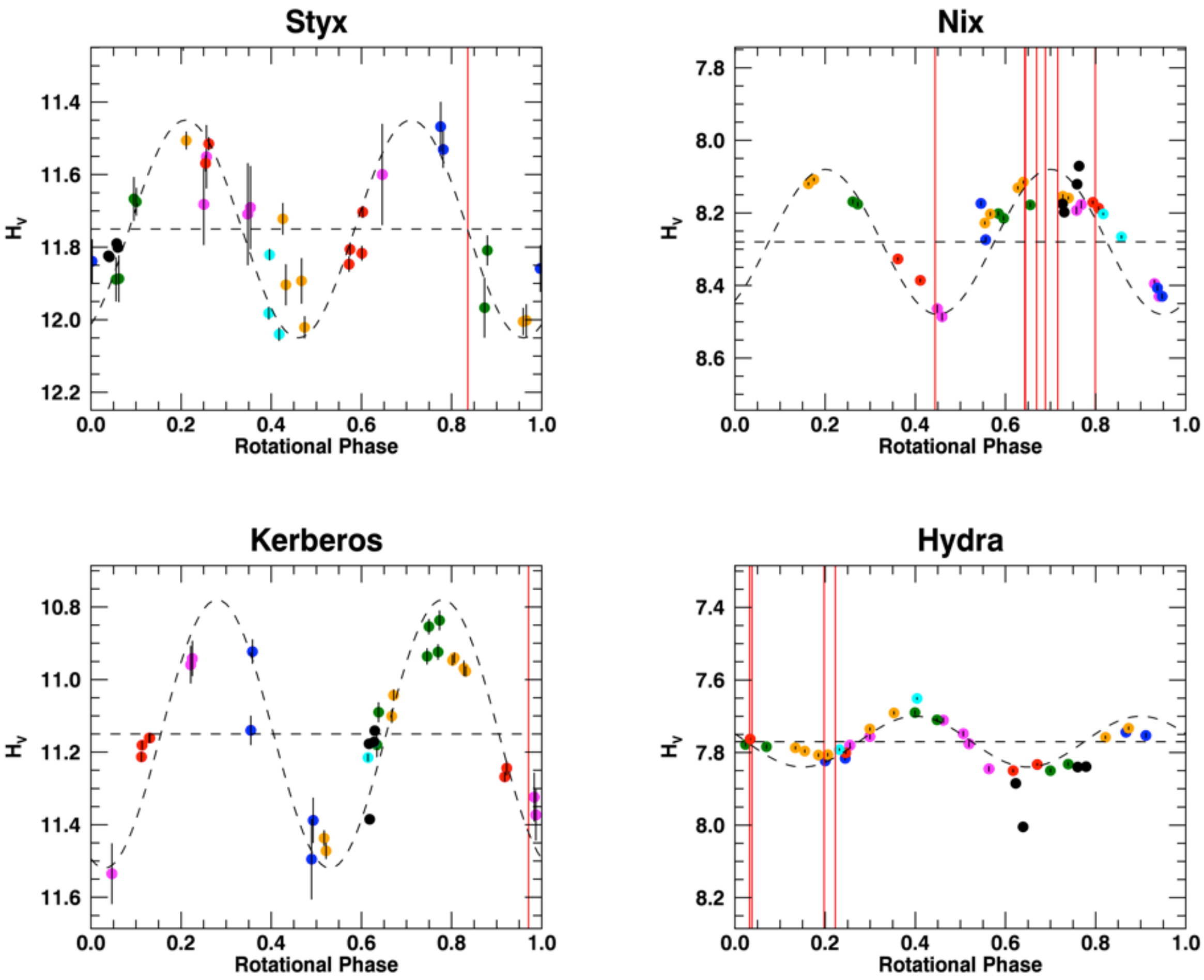

**Fig. 2: Rotational lightcurves for Pluto's small moons.** Systematic measurements of the brightnesses of Pluto's small satellites were obtained by LORRI during the approach to Pluto from May through early July 2015. $H_v$ refers to the total (i.e., integrated over the entire target) visible magnitude (V-band) referenced to a heliocentric distance of 1 AU, a spacecraft-to-target distance of 1 AU, and a solar phase angle of 0° (using a phase law of 0.04 mag/deg). Different colors are used for the seven different observing epochs (*12* and Table S4), and ±1σ error bars are shown for each measurement (some error bars are smaller than the plotting symbols). Three different algorithms were used to search for periodic variations in the data (*12*). The rotational periods derived from that analysis (Table 2) were then used to phase the brightness data,

producing the lightcurves displayed above. These double-peaked lightcurves presumably result from the rotation of elongated bodies, with the lightcurve amplitude determined by the variation in the cross sectional area presented to the observer, which depends on the body's shape and the angle between the rotational pole and the line-of-sight to the body. The rotational phase for all the resolved observations of the small satellites (Table 1) are indicated by the vertical red lines, although the angle between the observer and the rotational pole may be different for these observations compared to the earlier ones. The dashed curves are sinusoids with the best matched periods. The amplitudes for SNKH, respectively, are 0.30, 0.20, 0.37, and 0.07 mag. The dashed horizontal lines are the mean $H_v$ values, which are 11.75, 8.28, 11.15, and 7.77 mag for SNKH, respectively.

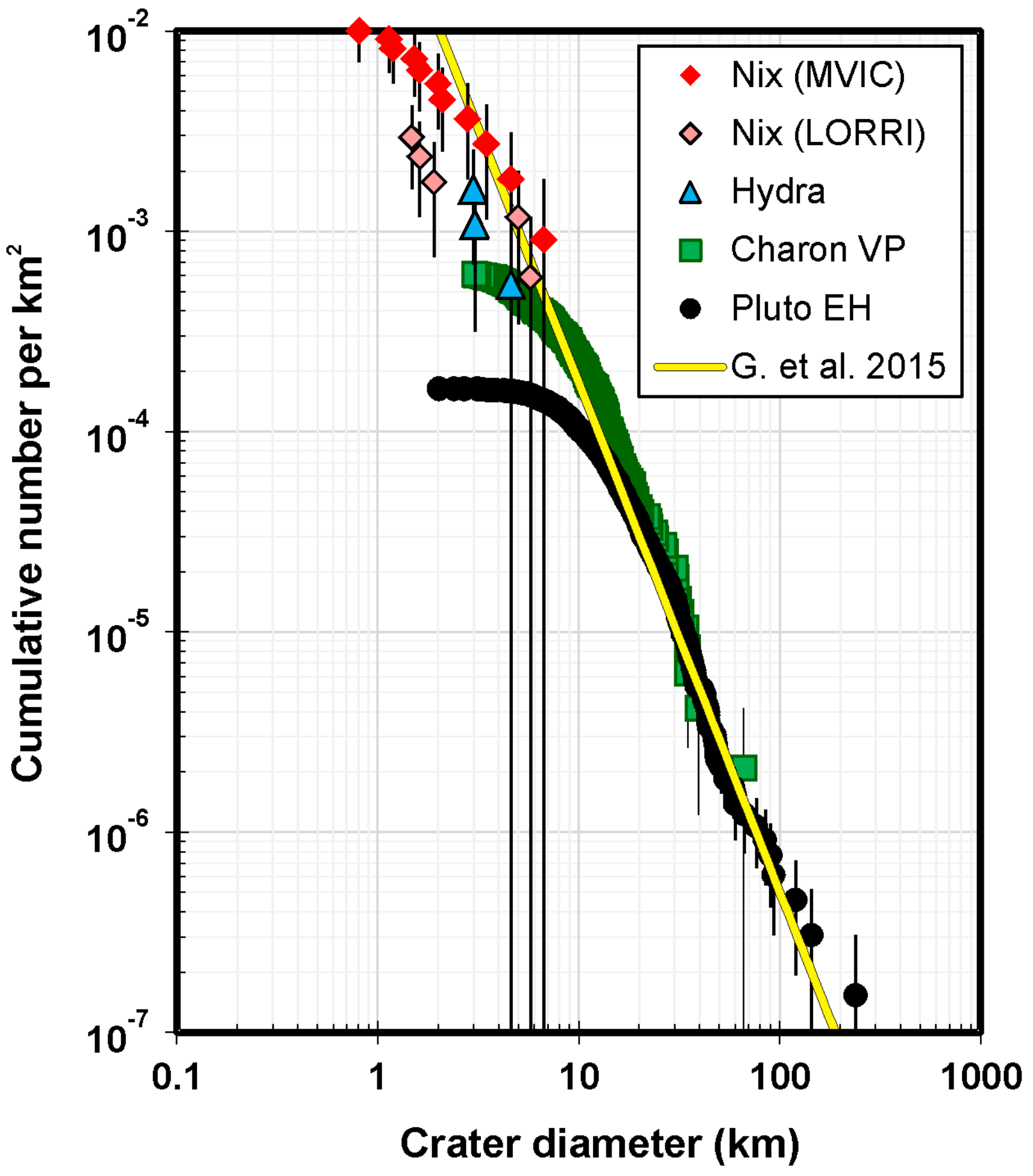

**Fig. 3: Cumulative crater size-frequency distributions for Nix, Hydra, Pluto's encounter hemisphere (EH), and Charon's Vulcan Planum (VP).** The curves for Pluto and Charon are from Moore et al. (*21*). Nix and Hydra crater sizes (Table S2) are scaled downward by a factor of 2.1 (appropriate for porous regolith-type material) to account for the difference in gravity between these small moons and Pluto; see the SOM for further details. Standard Poisson statistical errors ($\sqrt{N}$) are displayed. The phase angle for the "Nix (LORRI)" data (N_LEISA_LORRI_BEST; phase angle of 9.45°) was less ideal for topographic feature identification compared to the phase angle for the "Nix (MVIC)" observation (N_MPAN_CA; phase angle of 85.9°). Thus, the lower crater density for "Nix (LORRI)" versus "Nix (MVIC)" may be an artifact of the viewing and lighting geometry. The yellow line indicates the

Greenstreet et al. (*22*) prediction for the cumulative density of craters on Pluto's surface over 4 billion years for their "knee" model. Although not saturated in appearance, Nix and Hydra both exhibit slightly higher crater densities than Pluto and Charon, implying a surface age $\gtrsim$4 Ga (see text).

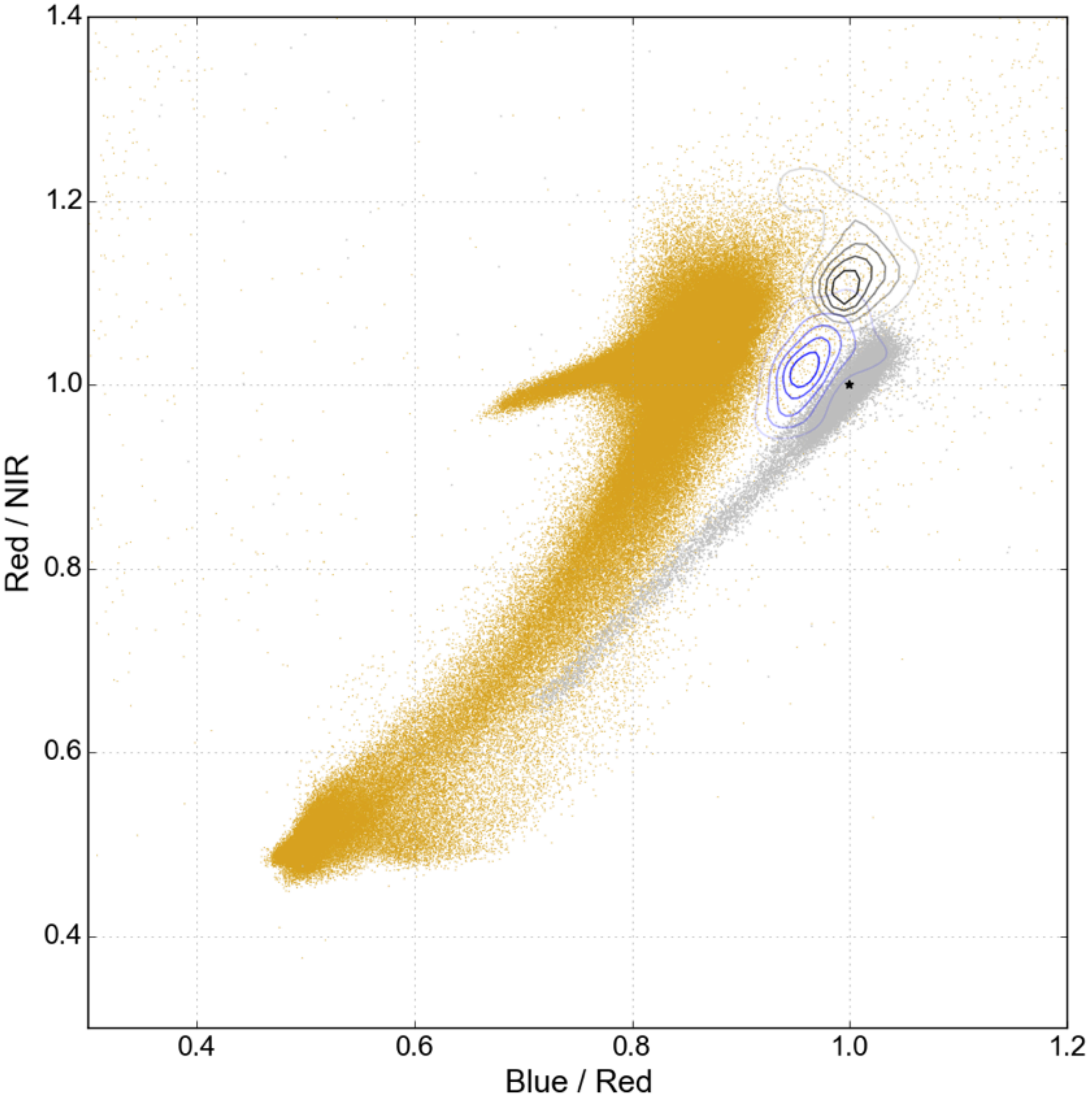

**Fig. 4. Color ratios for the surfaces of Pluto, Charon, Nix, and Hydra.** Blue/Red and Red/NIR color ratios derived from MVIC images are displayed (Blue = 400−550 nm, Red = 540−700 nm, NIR = 780−975 nm). Gold points are from Pluto's surface, and silver points are from Charon's surface. Blue contours show the distribution of colors on Nix's surface, and black contours show the distribution of colors on Hydra's surface. The normalized solar color is denoted by the star symbol (at coordinate [1,1] in the plot); surfaces redder than solar will be down and to the left of the star symbol, while bluer than solar regions will appear up and to the right. Pluto exhibits a diversity of colors over its surface (*24*). Charon has less color diversity than Pluto, and the range of its colors follows a mixing line (*24*). Nix (cyan contours) and Hydra (white contours) have nearly solar colors (i.e., grey color) that are distinct from either Pluto or Charon, with Hydra being slightly bluer than Nix.

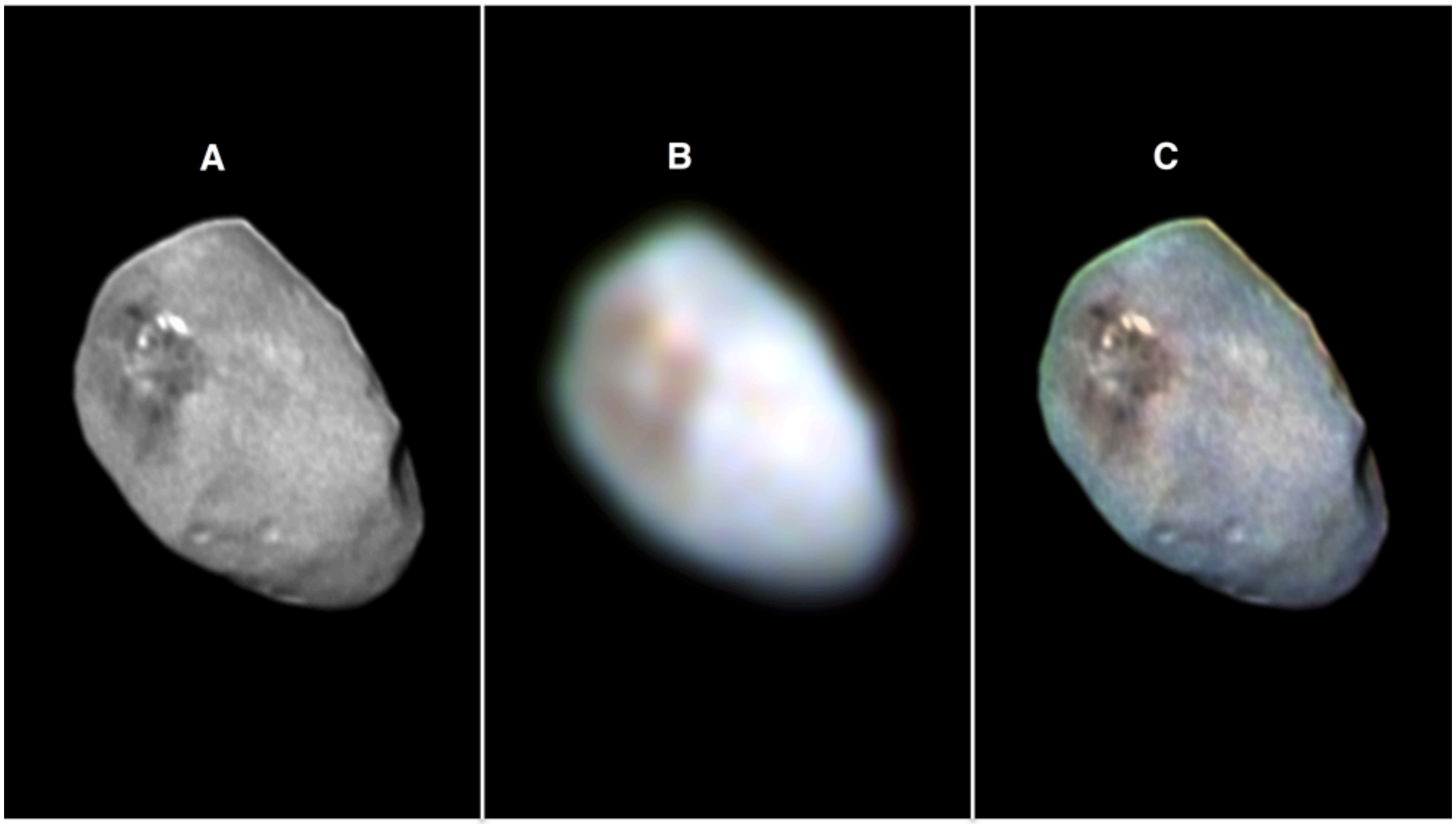

**Fig. 5. Color of Nix's surface.** (**A**) Panchromatic LORRI image of Nix taken from N_LEISA_LORRI_BEST (Table 1). (**B**) Enhanced MVIC color image of Nix taken from N_COLOR_BEST. (**C**) The LORRI image of Nix was colored using the data derived from the MVIC image. Most of Nix's surface is neutral (i.e., grey) in color, but the region near the largest impact crater is slightly redder than the rest of the surface. Celestial north is up and east is to the left for all images.

# Supplementary Materials for

## The Small Satellites of Pluto as Observed by *New Horizons*

H. A. Weaver,[1*] M. W. Buie,[2] B. J. Buratti,[3] W. M. Grundy,[4] T. R. Lauer,[5] C. B. Olkin,[2] A. H. Parker,[2] S. B. Porter,[2] M. R. Showalter,[6] J. R. Spencer,[2] S. A. Stern,[2] A. J. Verbiscer,[7] W. B. McKinnon,[8] J. M. Moore,[9] S. J. Robbins,[2] P. Schenk,[10] K. N. Singer,[2] O. S. Barnouin,[1] A. F. Cheng,[1] C. M. Ernst,[1] C. M. Lisse,[1] D. E. Jennings,[11] A. W. Lunsford,[11] D. C. Reuter,[11] D. P. Hamilton,[12] D. E. Kaufmann,[2] K. Ennico,[9] L. A. Young,[2] R. A. Beyer,[6,9] R. P. Binzel,[13] V. J. Bray,[14] A. L. Chaikin,[15] J. C. Cook,[2] D. P. Cruikshank,[9] C. M. Dalle Ore,[9] A. M. Earle,[14] G. R. Gladstone,[16] C. J. A. Howett,[2] I. R. Linscott,[17] F. Nimmo,[18] J. Wm. Parker,[2] S. Philippe,[19] S. Protopapa,[12] H. J. Reitsema,[2] B. Schmitt,[19] T. Stryk,[20] M. E. Summers,[21] C. C. C. Tsang,[2] H. H. B. Throop,[22] O. L. White,[9] A. M. Zangari[2]

[1]The Johns Hopkins University Applied Physics Laboratory, Laurel, MD 20723, USA.

[2]Southwest Research Institute, Boulder, CO 80302, USA.

[3]National Aeronautics and Space Administration (NASA) Jet Propulsion Laboratory, California Institute of Technology, Pasadena, CA 91109, USA.
[4]Lowell Observatory, Flagstaff, AZ 86001, USA.
[5]National Optical Astronomy Observatory, Tucson, AZ 26732 USA.
[6]Search for Extraterrestrial Intelligence Institute, Mountain View, CA 94043, USA.
[7]Department of Astronomy, University of Virginia, Charlottesville, VA 22904, USA.
[8]Department of Earth and Planetary Sciences, Washington University, St. Louis, MO 63130, USA.
[9]NASA Ames Research Center, Space Science Division, Moffett Field, CA 94035, USA.
[10]Lunar and Planetary Institute, Houston, TX 77058, USA.
[11]NASA Goddard Space Flight Center, Greenbelt, MD 20771, USA.
[12]Department of Astronomy, University of Maryland, College Park, MD 20742, USA.
[13]Massachusetts Institute of Technology, Cambridge, MA 02139, USA.
[14]University of Arizona, Tucson, AZ 85721, USA.
[15]Independent Science Writer, Arlington, VT USA.
[16]Southwest Research Institute, San Antonio, TX 78238, USA.
[17]Stanford University, Stanford, CA 94305, USA.
[18]University of California, Santa Cruz, CA 95064, USA.
[19]Université Grenoble Alpes, CNRS, IPAG, F-38000 Grenoble, France.
[20]Roane State Community College, Oak Ridge, TN 37830, USA.
[21]George Mason University, Fairfax, VA 22030, USA.
[22]Planetary Science Institute, Tucson, AZ 85719, USA.

correspondence to: hal.weaver@jhuapl.edu

**This PDF file includes:**

Materials and Methods
Supplementary Text
Tables S1 – S4
Figs. S1 – S14
Supplementary movie (mp4 format)

## Materials and Methods

Image Processing Method

The processing of the LORRI images of the small moons was optimized to recover the highest spatial resolution information captured by the instrument. The LORRI point spread function (PSF) is slightly undersampled by the detector pixels, so the raw LORRI images are aliased. For most of the moons, however, multiple images were taken during the observational sequences, and random pointing variations effectively "dithered" the moons over the LORRI field of view. We adopted a Fourier-based reconstruction technique (*25*) and used the sub-pixel shifts over a given set of images to construct a Nyquist-sampled "superimage", capturing the complete structural information available over the complete sequence. The superimage is generated with a pixel scale twice as fine as the native LORRI sampling. This image is then deconvolved using the Lucy-Richardson (*26,27*) algorithm to correct for the LORRI PSF. The input PSFs were generated from LORRI images of open star clusters and are likewise double-sampled. The PSFs were further generated in a 3×3 partitioning of the LORRI field to account for spatial variability of the PSF. In some cases, the final deconvolved image was up-sampled to an even finer level of sub-sampling for clarity of visualization. Since the super-image was Nyquist-sampled, this procedure is mathematically well-justified. Figure S1 graphically shows the progression of image processing steps.

Rotation Rate Determinations

The rotation rates of the small moons were determined by performing a harmonic analysis of brightness data taken over several months during the *NH* spacecraft's approach to Pluto. Most of these brightness data were acquired as part of a systematic "hazard mitigation" campaign to search for new satellites or dust in the Pluto system (impacts from even mm-size dust grains could potentially damage the spacecraft). Those observations were acquired at seven different epochs (all in 2015): May 11 (144 images), May 29 (144 images), June 5 (216 images), June 15 (384 images), June 23 (72 images), June 26 (48 images), and July 1 (48 images). Each epoch itself was comprised of either 2, 3, or 4 "visits", separated in time by several hours to a day depending on the epoch. To improve our sensitivity to faint objects, we typically created composite images by combining many (36, 48, or 96) 10s exposures, which means we would not be able to detect temporal variations shorter than ~6-16 minutes. Although not shown below, we expanded the period search for Nix and Hydra down to 0.01 days (14.4 min), but no significant peaks were identified for rotational periods between 0.01 and 0.1 days. The total time span of the hazard program was 51 days, but we also added data for Nix, Hydra, and Kerberos from earlier "opnav" observations (starting on April 5 for Nix and Hydra; starting on April 15 for Kerberos. Thus, our sampling period easily covered the longest orbital period of Pluto's satellites (38.2 days for Hydra).

Three different period search techniques were employed: one developed by author Showalter (our primary method) and two using online routines hosted at the NASA Exoplanet Science Institute (used as independent verification).

Four plots illustrating Showalter's technique are shown below (Fig. S2). The figure caption explains what is being plotted and summarizes the results. For each of the deepest minima for each moon (which we take as the rotation period), the error is determined by the half-width at half-minimum of each of those minima. The median of the other root-mean-square (RMS) values in the plot defines the baseline. These errors are listed in Table 2.

We used the Lomb-Scargle (*28*) and Plavchan (*29*) online time series analysis programs from the NASA Exoplanet Science Institute to verify Showalter's results. The Lomb-Scargle (LS) technique provides an approximation for the Fourier transform decomposition of data that are unevenly spaced in time, and it generally performs well when the time series is comprised of combinations of sinusoids (i.e., it is basically a more generalized version of Showalter's technique). The Plavchan (Plav) technique is essentially a phase dispersion minimization algorithm that identifies periods with coherently phased time-series curves, otherwise making no assumptions about the underlying shape of the periodic signal (i.e., does not assume sinusoids). For all 4 moons, the periods determined by the LS and Plav routines agree with Showalter's results to within Showalter's ±1σ error bars, although the period found by Showalter for Styx was the second ranked period by the LS and Plav routines (i.e., one other period had larger peaks in the periodograms). We have adopted the rotational periods derived by Showalter here (Table 2).

Rotation Pole and Shape Determinations

Adopting the rotational periods discussed above for each of Pluto's small moons, author Porter constructed a model using 3-dimensional ellipsoid representations for each moon. The shape of each moon and the locations of the rotational poles were determined by minimizing the differences between the observed and model shapes and brightnesses; multiple Monte Carlo model runs were performed to best fit both the lightcurve and resolved images of the satellites. In both cases, the satellite's apparent shape at the time of the observation was directly rendered at the pixel scale of the LORRI camera. For resolved observations, this was then convolved with the LORRI PSF to produce a synthetic image. For unresolved observations, the image was summed and the resulting brightness converted to instrumental magnitude. The results for the derived pole locations are shown in Fig. S4. The pole locations and size determinations are provided in Table 2.

Gravity-scaled crater sizes for Nix and Hydra

To facilitate the comparison with crater statistics on Pluto, we estimated an equivalent, gravity-scaled size for the craters on Nix and Hydra. The average diameter (from the cube root of the product of the three dimensions given in Table 2), radius (*R*), and *GM* (4; *GM* is the product of the Newton's gravitation constant and the body's mass) used to calculate surface gravity (*g*) are given in Table S1. The calculated surface gravities for both bodies are close to 0.01 m $s^{-2}$, roughly 70 times lower than Pluto's. Schmidt-Holsapple-Housen impactor scaling predicts that the crater diameter for a given size impactor scales with $g^{\sim 0.216}$ in the gravity regime (*30*) for cold, non-porous ice, or $g^{\sim 0.170}$ for porous regolith-like material. This yields gravity scaling factors of ~2.1 or ~2.6 for cold ice or regolith, respectively; that is, the same impactor would make a 2.1-2.6 times smaller crater on Pluto than on Nix or Hydra. These small moons are likely rubble piles, so we choose regolith scaling as the most reasonable material. A comparison of the crater densities on Nix, Hydra, Pluto, and Charon using this scaling is shown in Fig. 3. Fig. S5 shows this comparison using cold ice scaling, and Fig. S6 shows the comparison with no scaling.

Slightly lower impact speeds on Nix and Hydra compared with Pluto and Charon have a very modest effect on crater size, one that is easily subsumed into that due to the uncertainty in the masses and gravities on the two small bodies, so we ignore it here. However, while gravitational focusing is essentially irrelevant for Nix and Hydra, is it not for Pluto. This is a velocity dependent quantity, and impact speeds onto Pluto and Charon cover a broad range (*22*). For a mean impact speed onto Pluto of ~2 km/s, the crater densities on Nix and Hydra may reduced by a multiplicative factor of 2/3 in Figs. 3 and S5.

Crater identification on Nix and Hydra:

Craters were mapped by three authors: H. Weaver, S. Robbins, and K. Singer. The identifications were ranked as “certain”, “likely”, or “tentative” and an average crater diameter was computed. The cumulative size-frequency distribution in Figure 3 contains the “certain” and “likely” craters identified by 2 or 3 researchers. Table S2 and Figures S13-S15 display these craters. Other dark-bright patterns on the surfaces of Nix and Hydra may be indicating a topographic depression that is an impact crater, but degradation or softening of the features, or lack of adequate image resolution, prevent positive identification. Thus, the small moons results plotted in Figure 3 may represent a lower bound on crater densities. Even given the uncertainties in crater identification and in the area measured (see Table S3), the crater densities on both Nix and Hydra are quite high, indicating relatively old surfaces.

## Supplementary Text

Montage of Resolved LORRI Images of Nix

Four resolved LORRI images of Nix are displayed in Figure S7.

I/F contours for the best images of Styx, Nix, Kerberos, and Hydra

The best images of SNKH were converted to reflectance (I/F) images with I/F contours drawn on top of the images: Fig. S8 shows Styx, S9 shows Nix, S10 shows Kerberos, and S11 shows Hydra.

Images of Nix and Hydra showing crater locations

Fig. S12 shows the highest resolution image of Nix with the locations of craters marked by labels. Fig. S13 shows a high phase angle image of Nix with the locations of craters marked by labels. Fig. S14 shows the best image of Hydra with the locations of craters marked by labels. In all cases, the labels refer to the crater IDs in Table S2.

Caption for "Spinning Moons" movie

Most inner moons in the Solar System keep one face pointed toward their central planet; this animation shows that certainly is *not* the case with the small moons of Pluto. Whereas Pluto and Charon rotate synchronously, always keeping the same face toward the other object as they revolve around their common center of mass (at the "barycenter" of the Pluto system, at the center of the frame), the small moons behave like spinning tops, rotating about their axes much faster than they revolve around the system barycenter. In this movie, Pluto is shown near the center of the frame with, in order, from smaller to wider orbits: Charon, Styx, Nix, Kerberos and Hydra. Nix rotates in the opposite sense to its orbital motion around the barycenter (i.e., Nix displays "retrograde" rotation). Although the relative spin rates and orbital periods are depicted accurately in this animation, no attempt has been made to depict the rotational pole locations of the small moons. The pole locations for the small moons are clustered near their orbital planes (i.e., nearly orthogonal to the mutual rotational pole locations for Pluto and Charon). See the main paper and the other Supplemental Online material, for further discussion.

**Table S1. Gravity scaling of crater sizes.** For Nix and Hydra, the second column is the average diameter for a sphere having the same volume as a body with the dimensions listed in Table 1; the third column is the product of Newton's gravitation constant and the object's mass; the fourth column is the surface gravity; the fifth and sixth columns are the scaling factors used to convert raw crater diameters to their equivalent values on Pluto, assuming the body has the density of either cold ice or regolith, respectively.

| Value: | Average Diameter [km] | $GM^*$ [$km^3 s^{-2}$] | $g$ [$m s^{-2}$] | Gravity Scaling Factor - Cold Ice | Gravity Scaling Factor - Regolith |
|---|---|---|---|---|---|
| Nix | 39 | 0.0030 | 0.008 | 2.6 | 2.1 |
| Hydra | 42 | 0.0032 | 0.007 | 2.7 | 2.2 |

*From Brozović et al. (*4*).

**Table S2. Crater Identifications.** For the three images listed along the top row, we list the craters identified by 2 or 3 independent researchers and their size estimates.

| N_MPAN_CA | | N_LEISA_LORRI_BEST | | H_LORRI_BEST | |
|---|---|---|---|---|---|
| Crater ID | Consensus Diameter [km]* | Crater ID | Consensus Diameter [km]* | Crater ID | Consensus Diameter [km]* |
| 1 | 14.1 ± 1.0 | 1 | 12.1 ± 3.4 | 1 | 9.7 ± 1.5 |
| 2 | 9.7 ± 1.0 | 2 | 10.5 ± 1.3 | 2 | 6.4 ± 0.2 |
| 3 | 7.3 ± 1.0 | 3 | 3.4 ± 1.0 | 3 | 6.3 ± 0.3 |
| 4 | 5.9 ± 0.3 | 4 | 3.1 ± 1.2 | | |
| 5 | 3.2 ± 0.5 | 5 | 4.0 ± 0.5 | | |
| 6 | 2.4 ± 0.1 | | | | |
| 7 | 2.5 ± 0.4 | | | | |
| 8 | 1.7 ± 0.1 | | | | |
| 9 | 4.4 ± 0.5 | | | | |
| 10 | 3.4 ± 0.6 | | | | |
| 11 | 4.2 ± 1.5 | | | | |

*Consensus crater diameters are given to a tenth of a kilometer (km) to show the average consensus value of the diameter, *not* to indicate that the values are known to this accuracy. Standard deviation of the measurements of 2-3 researchers for each crater is shown and represents the differences in selected rim positions. The estimated measurement error on crater diameters, if the rim location were known exactly, would still be ~1 pixel (see Table S3 for the pixel scales).

**Table S3. Areas for crater density estimates.**

| Observation Name | Pixel Scale [km/px] | Projected Mapping Area [$km^2$]* | Surface Area of Ellipsoid [$km^2$] † | Percent of Ellipsoid Visible‡ | Ellipsoidal Mapping Area [$km^2$]§ | Surface Area used [$km^2$]◊ |
|---|---|---|---|---|---|---|
| N_MPAN_CA | 0.45 | 1000 | 4800 | 25% | 1200 | 1100 |
| N_LEISA_LORRI _BEST | 0.31 | 1200 | 4800 | 45% | 2200 | 1700 |
| H_LORRI_BEST | 1.14 | 1200 | 6200 | 40% | 2000 | 1600 |

*These are the areas of the illuminated region of the body computed by multiplying the area in $pixels^2$ by the square of the pixel scale. For N_MPAN_CA, the area was increased by 20% because some of the measured craters extended past the terminator.

†These are the total surface areas, using the ellipsoidal axes dimensions from Table 2.

‡Approximate percent of each moon that is visible.

§Column 4 multiplied by column 5.

◊Average of projected and ellipsoidal mapping area estimates. This area is used in Fig. 3 to calculate crater densities. A formal error associated with the area uncertainty was not included in Fig. 3, but even using the most extreme values for the surface areas (column 3 versus column 6) doesn't change the crater densities by more than a factor of 2. Thus, the conclusion that Nix and Hydra have high crater densities, implying surface ages $\gtrsim$ 4 Ga, is robust.

**Table S4. Color scheme for Fig. 2 plot symbols.** A different color is used for each of the 7 different observing epochs (*12*). This table shows the correspondence between the plot symbol color and the observing epoch.

| **Color** | **Epoch (2015)** |
|---|---|
| Magenta | May 11 |
| Blue | May 29 |
| Green | June 5 |
| Orange | June 15 |
| Red | June 23 |
| Cyan | June 26 |
| Black | July 1 |

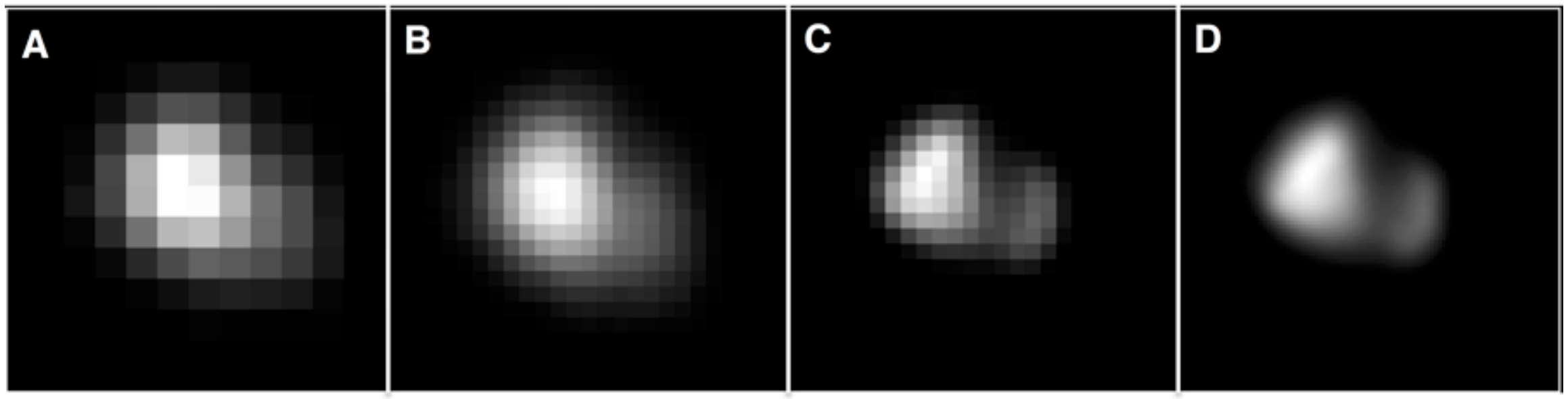

**Fig. S1. Processing of Images.** Demonstration of the typical image processing steps performed on the images of the small satellites, as applied to the LORRI Kerberos observations. **(A)** A single calibrated LORRI image, one of four taken, is shown at the far left. **(B)** The interlaced and Nyquist-sampled "superimage" generated by combining the four calibrated images − the pixel scale is twice as fine as the native LORRI scale. **(C)** The superimage after applying Lucy-Richardson deconvolution. **(D)** The deconvolved image up-sampled by an additional factor of 4 for a final scale 8 times finer than the native LORRI scale. This latter step removes the pixelated appearance of the previous image for improved clarity, and is mathematically justified since the superimage is Nyquist-sampled.

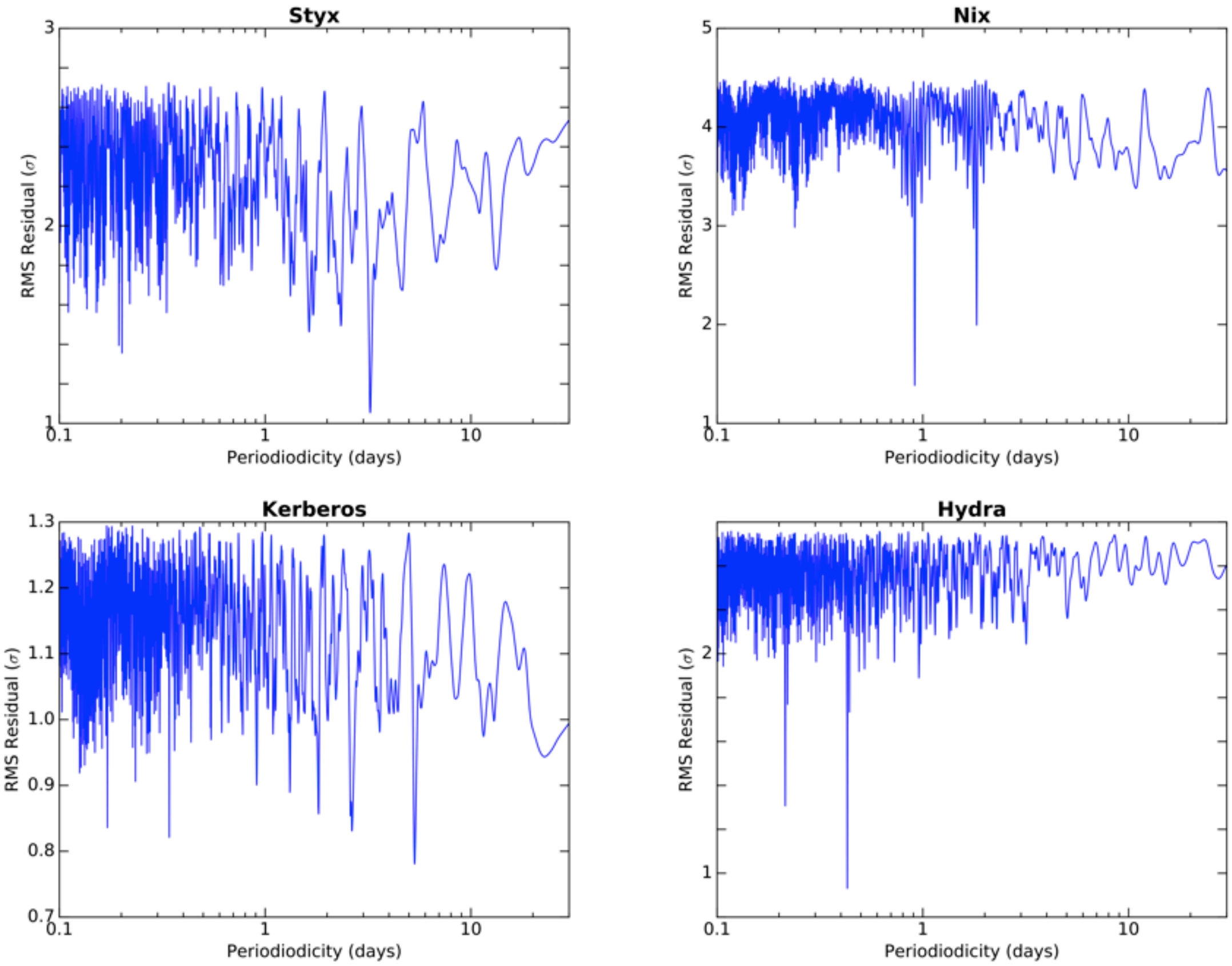

**Fig. S2: Harmonic time series analysis of the brightness (lightcurve) data for Pluto's small moons.** Each point in each plot is created by picking a period and then fitting a sinusoid of that period, plus its first harmonic, to the data. The difference between the model and the data provide "residuals", which will be minimized for the best-fit period. Typically one particular period, and possibly either half or twice that period, will stand out with the smallest residuals relative to other potential periods. Such is the case for Styx (3.24 day period), Nix (1.829 day double-peaked period, with the smallest residual at 0.915 day), and Hydra (0.429 day period). The case for Kerberos is more complicated, but phasing the data using a period of 5.31 days (where the minimum residual lies) gives an acceptable fit.

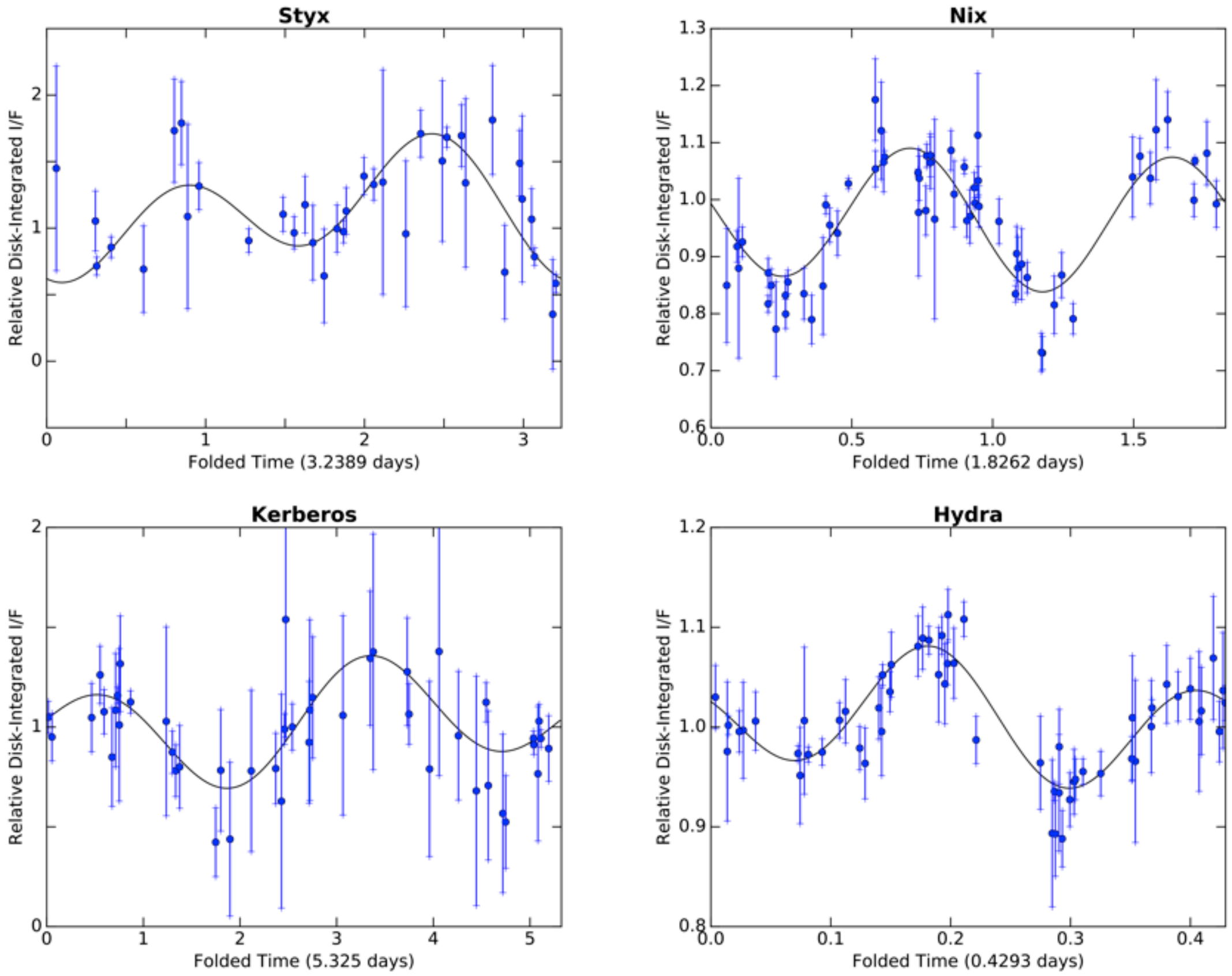

**Fig. S3: Lightcurves for Pluto's small moons.** In the plots above, the brightness data from each moon have been phased to the rotational period determined from the Showalter analysis technique. These results are similar to those shown in Fig. 2.

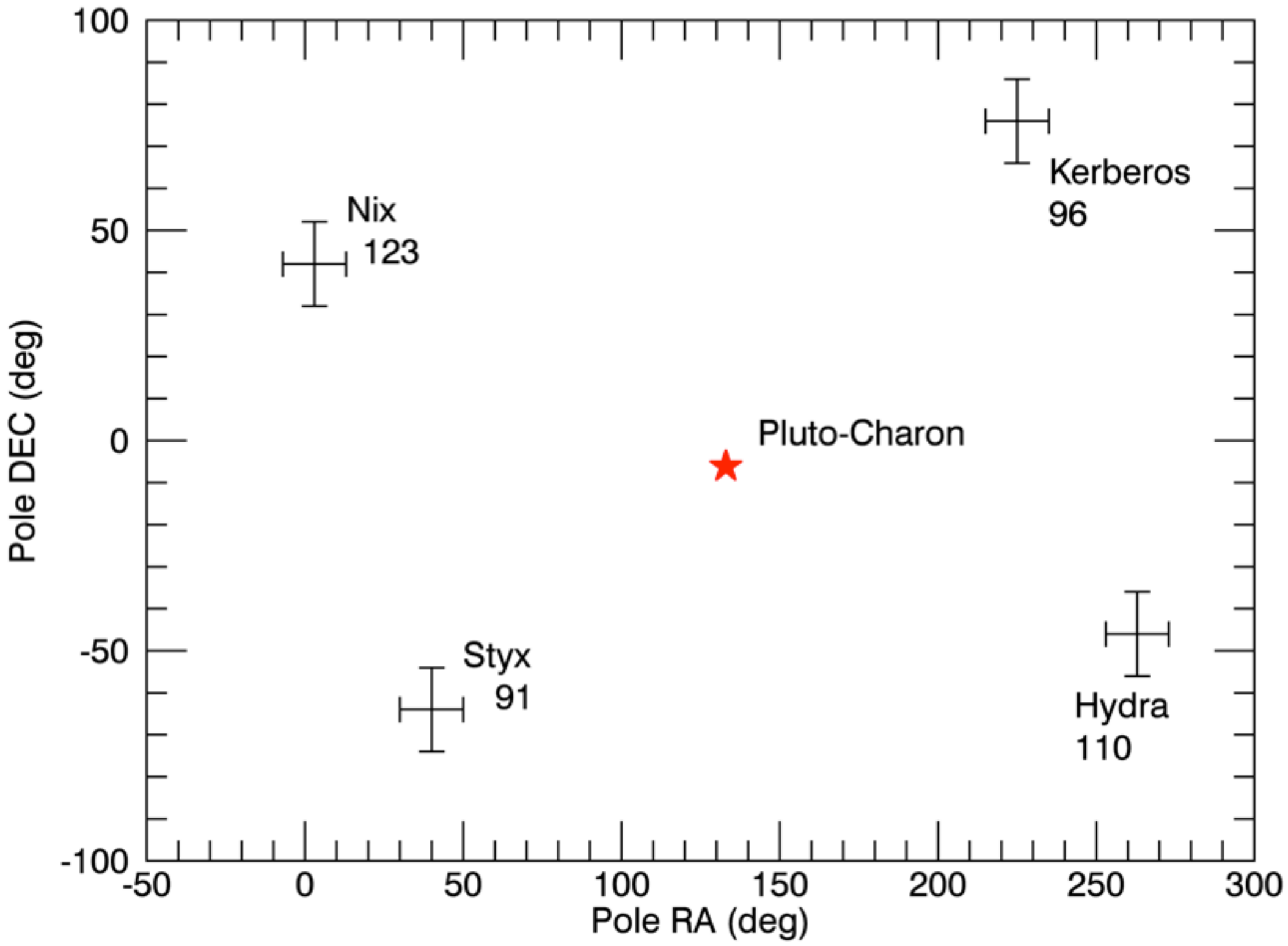

**Fig. S4. Rotation pole locations for Pluto's small moons.** The locations of the north poles (i.e., directions determined using the right-hand-rule with the hand curling around in the spin direction and the thumb denoting the north pole) for Pluto's small moons are plotted on a grid of celestial coordinates in the standard J2000 frame (RA is right ascension; DEC is declination). The pole locations were determined using the analysis method described above. The direction of Pluto's and Charon's north pole (which are identical) is also plotted (red star symbol). The angle (in degrees) between the Pluto-Charon pole location and the pole location for each of the small moons (i.e., the obliquity) is listed near the satellite name; these values are uncertain by ±10°. When that angle is larger than 90°, the moon rotates in the opposite direction from its orbital motion around Pluto (i.e., the rotation is retrograde, rather than prograde). Nominally, all the small moons have retrograde rotation, but Nix is the only one significantly so (i.e., retrograde with greater than 1σ confidence). All four moons have rotation poles clustered near their orbital planes (i.e., near the Pluto-Charon orbital plane).

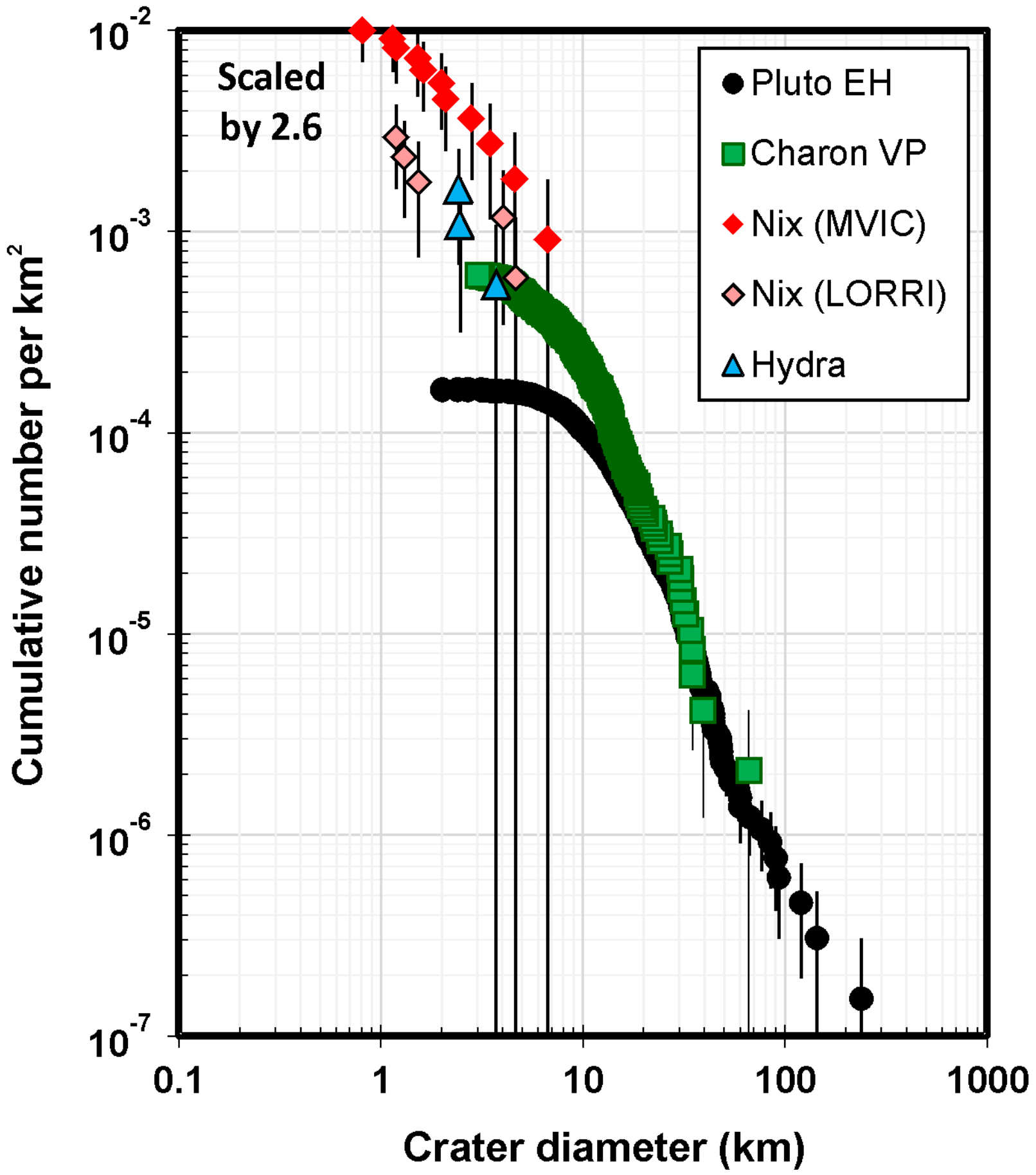

**Fig. S5. Cumulative crater size-frequency distributions for Nix, Hydra, Pluto's encounter hemisphere (EH), and Charon's Vulcan Planum (VP).** As in figure 3, but assuming solid ice scaling factors for Nix and Hydra.

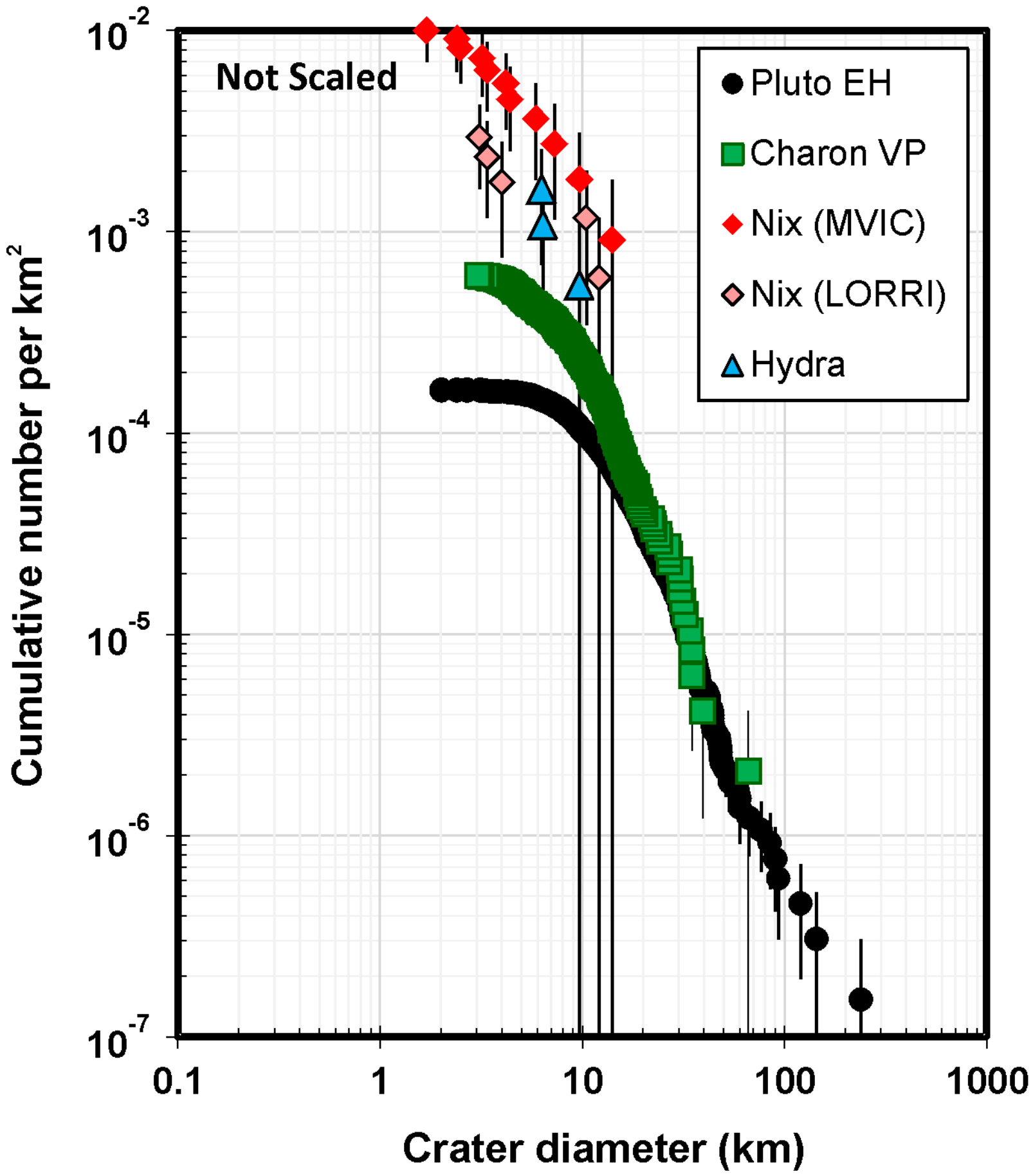

**Fig. S6. Cumulative crater size-frequency distributions for Nix, Hydra, Pluto's encounter hemisphere (EH), and Charon's Vulcan Planum (VP).** The curves for Pluto and Charon are from Moore et al. (*20*). As in figure 3 and S5, but Nix and Hydra crater sizes have *not* been scaled to account for the difference in gravity between these small moons and Pluto (i.e., the raw crater densities for Nix and Hydra are plotted).

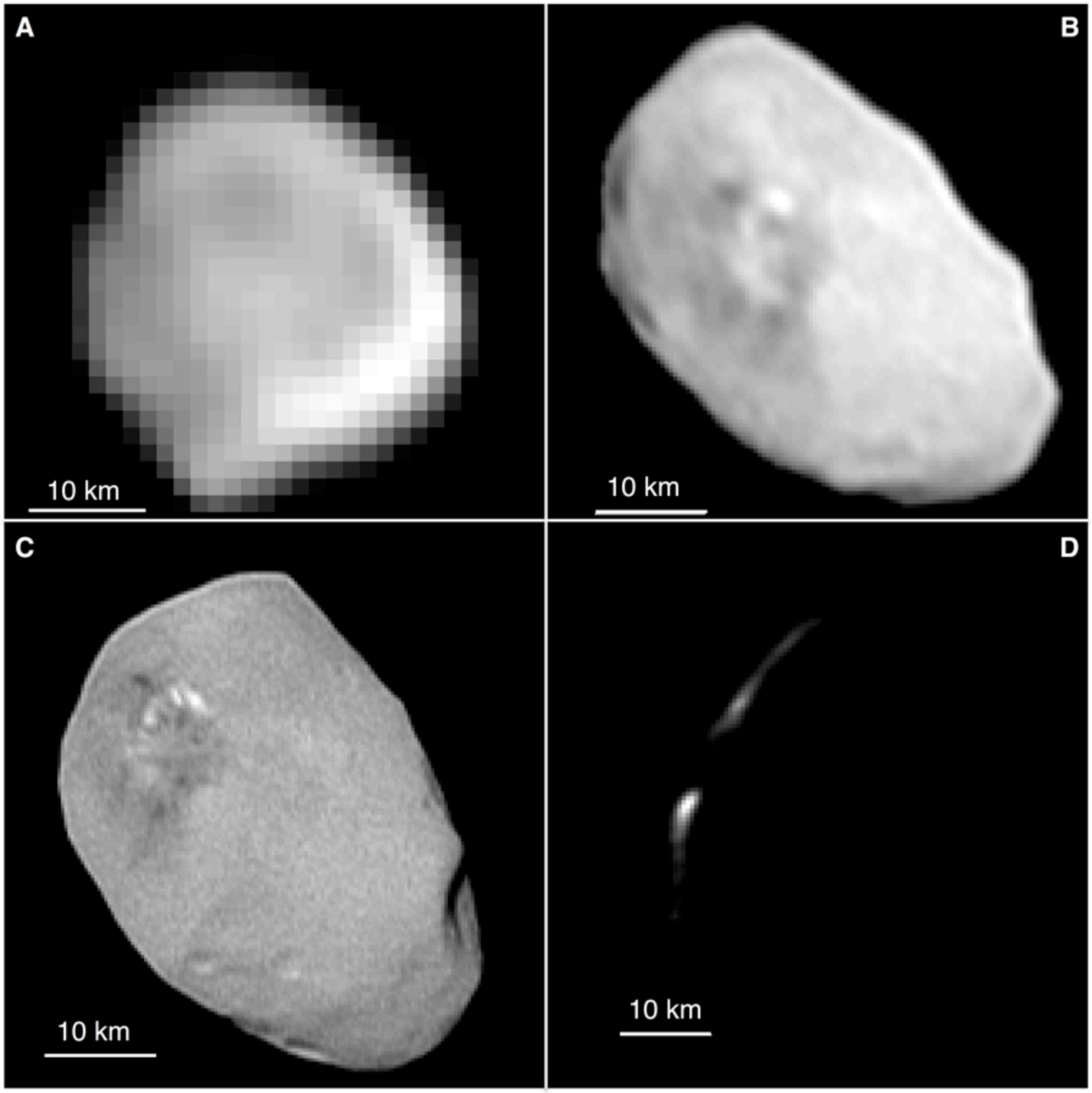

**Fig. S7: Montage of four resolved Nix images taken by LORRI.** All images have been deconvolved and are displayed with celestial north up and east to the left. **(A)** Image from N_LORRI_APPR_1D2 (Table 1), taken at a phase angle of 13.4° and a resolution of 2.92 km/pixel. (**B**) Image from N_LORRI_BACKUP, taken at a phase angle of 8.34° and a resolution of 0.76 km/pixel. (**C**) Image from N_LEISA_LORRI_BEST, taken at a phase angle of 9.45° and a resolution of 0.30 km/pixel. (**D**) Image from N_DEP_SOONEST, taken at a phase angle of 158° and a resolution of 0.93 km/pixel.

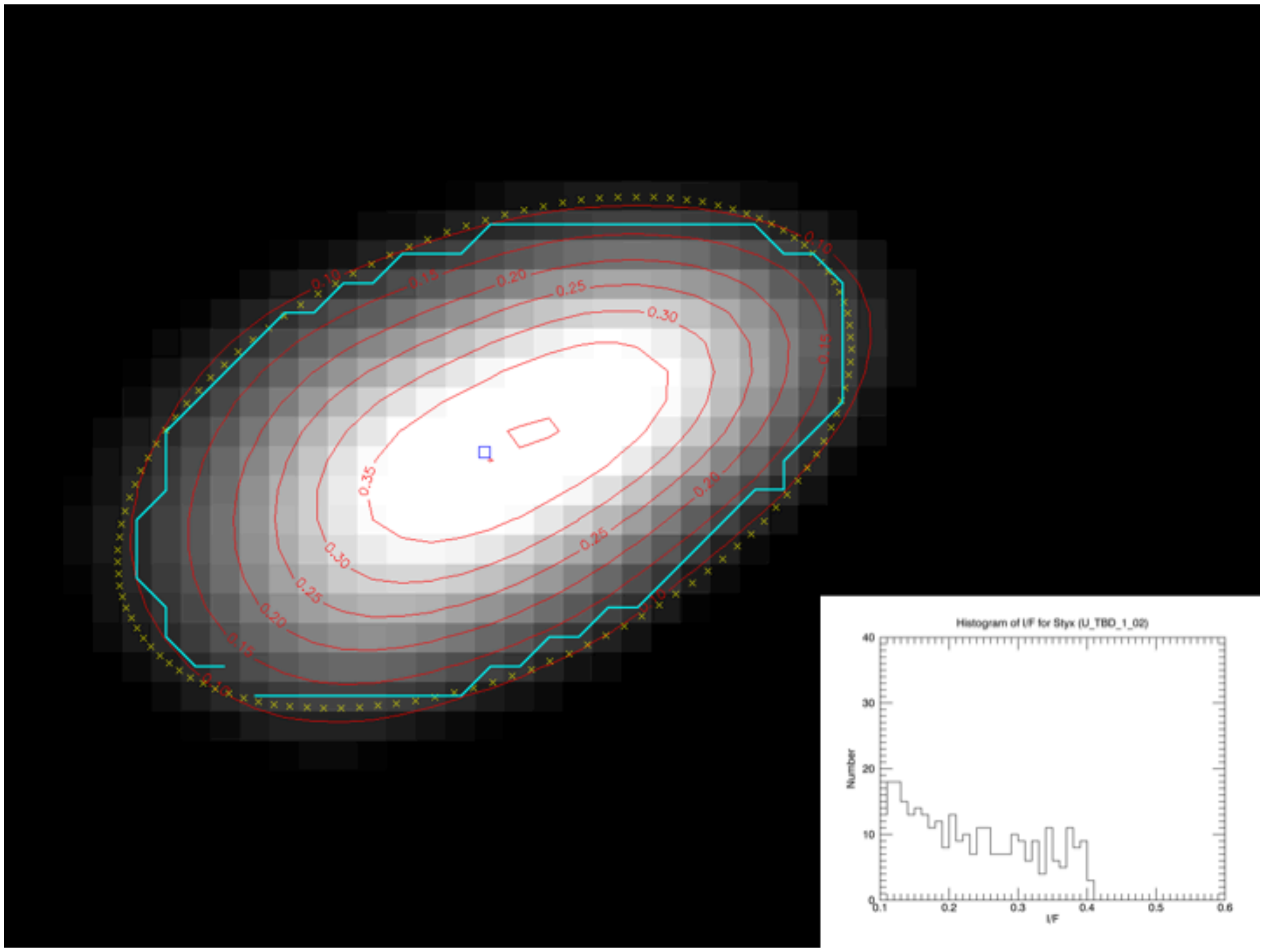

**Fig. S8: I/F image of Styx.** The best image of Styx (deconvolved and supersampled by a factor of 8) has been converted from instrumental units to reflectance (I/F) values. I/F contours have been drawn on top of the image. The best-fit ellipsoid (10.4 km × 5.8 km) using an I/F threshold of 0.10 is depicted by yellow x symbols. The boundary determined by I/F ≥ 0.10 is indicated by the cyan line. The boundary perimeter is ~27 km. The boundary surface area is ~47 $km^2$. The small blue box shows the center of the ellipse and the boundary. The inset figure at the bottom right is a histogram of the I/F values; the peak I/F value is 0.40. Celestial north is up and east is to the left.

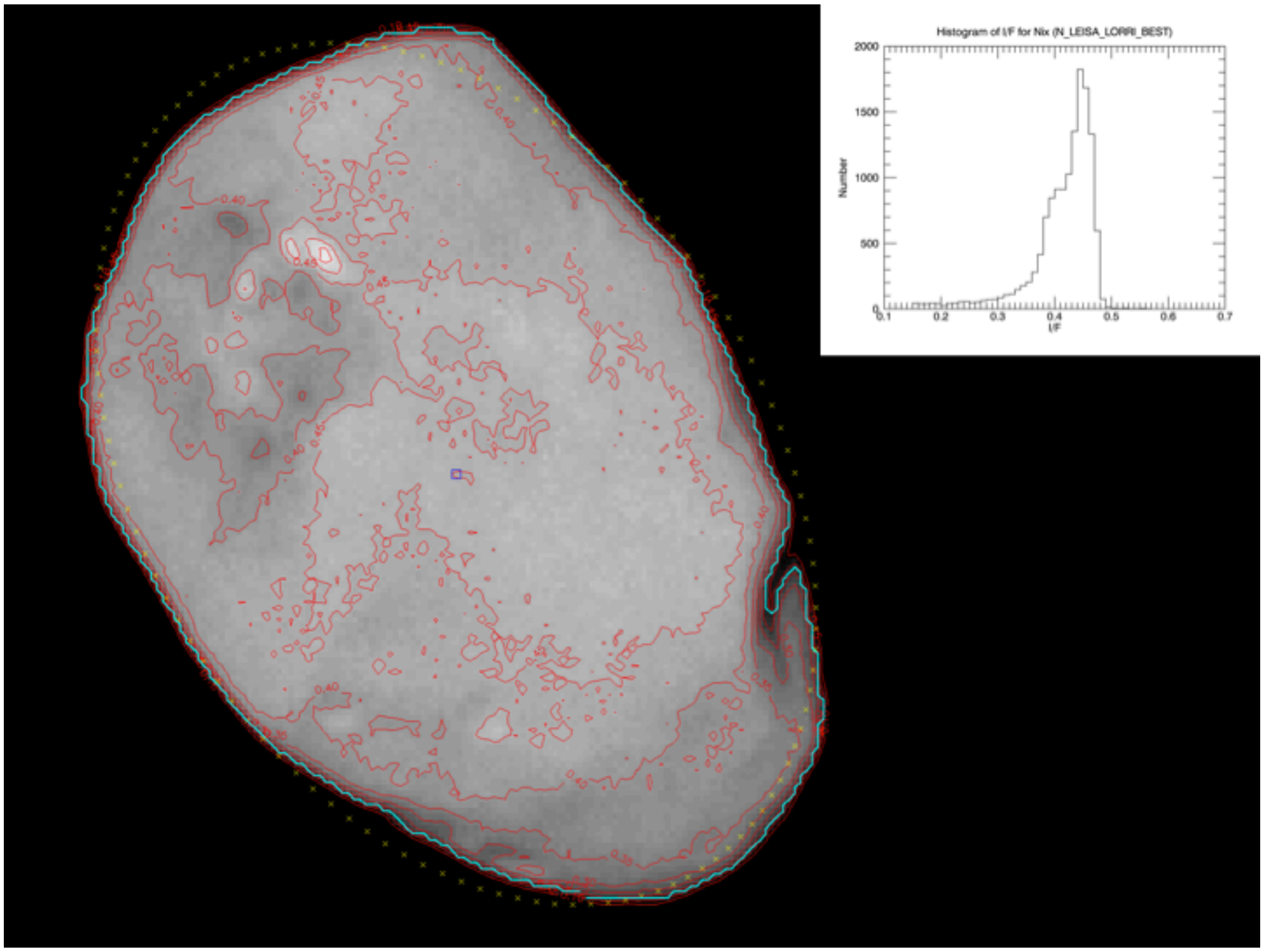

**Fig. S9: I/F image of Nix.** The best image of Nix (from N_LEISA_LORRI_BEST; raw pixel version) has been converted from instrumental units to reflectance (I/F) values. I/F contours have been drawn on top of the image. The best-fit ellipsoid (48.5 km × 32.7 km) using an I/F threshold of 0.15 is depicted by yellow x symbols. The boundary determined by I/F ≥ 0.15 is indicated by the cyan line. The boundary perimeter is ~143 km. The boundary surface area is ~1235 km$^2$. The small blue box shows the center of the ellipse and the boundary. The inset figure at the top right is a histogram of the I/F values; the peak I/F value is 0.57. Celestial north is up and east is to the left.

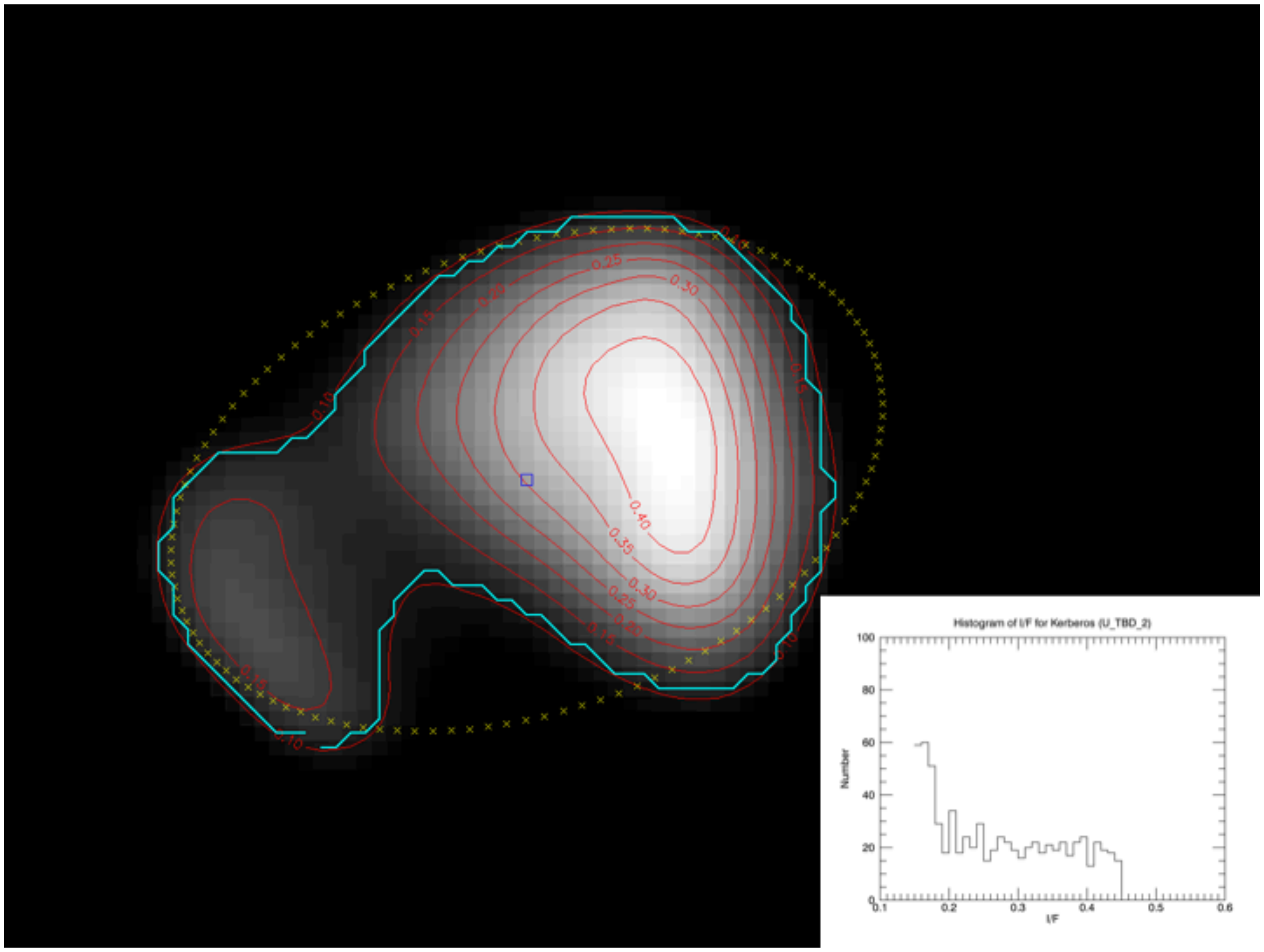

**Fig. S10: I/F image of Kerberos.** The best image of Kerberos (deconvolved and supersampled by a factor of 8) has been converted from instrumental units to reflectance (I/F) values. I/F contours have been drawn on top of the image. The best-fit ellipsoid (12.3 km × 7.73 km) using an I/F threshold of 0.10 is depicted by yellow x symbols. The boundary determined by I/F ≥ 0.10 is indicated by the cyan line. The boundary determined by I/F ≥ 0.10 is indicated by the cyan line. The boundary perimeter is
~37 km. The boundary area is ~66 $km^2$. The small blue box shows the center of the ellipse and the boundary. The inset figure at the bottom right is a histogram of the I/F values; the peak I/F value is 0.45. Celestial north is up and east is to the left.

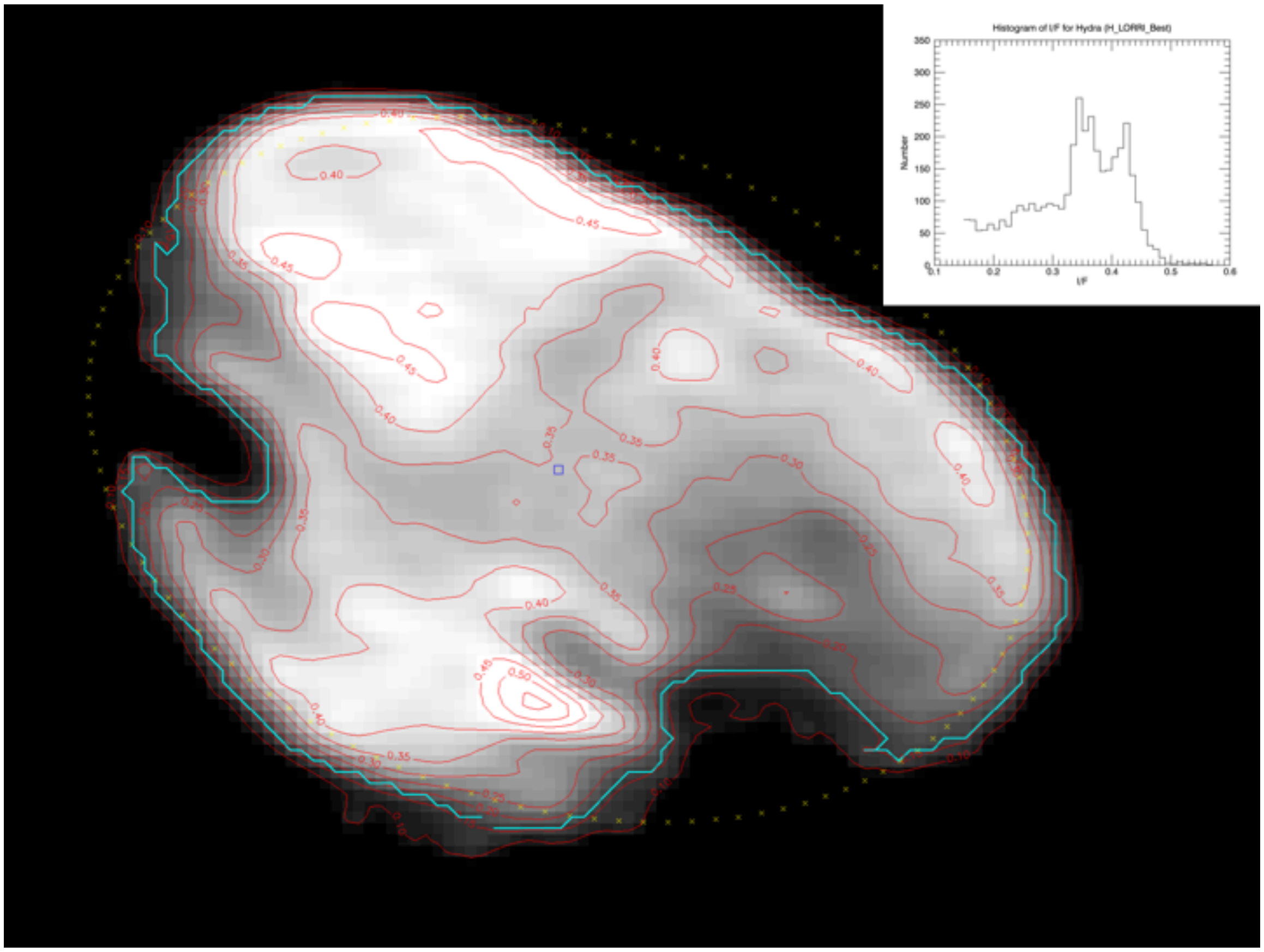

**Fig. S11: I/F image of Hydra.** The best image of Hydra (deconvolved and supersampled by a factor of 2) has been converted from instrumental units to reflectance (I/F) values. I/F contours have been drawn on top of the image. The best-fit ellipsoid (48.8 km × 33.9 km) using an I/F threshold of 0.10 is depicted by yellow x symbols. The boundary determined by I/F ≥ 0.10 is indicated by the cyan line. The boundary perimeter is
~165 km, but some of that is following indentations we identify as craters on the surface. The boundary surface area is ~1210 km$^2$. The small blue box shows the center of the ellipse and the boundary. The inset figure at the top right is a histogram of the I/F values; the peak I/F value is 0.56. Celestial north is up and east is to the left.

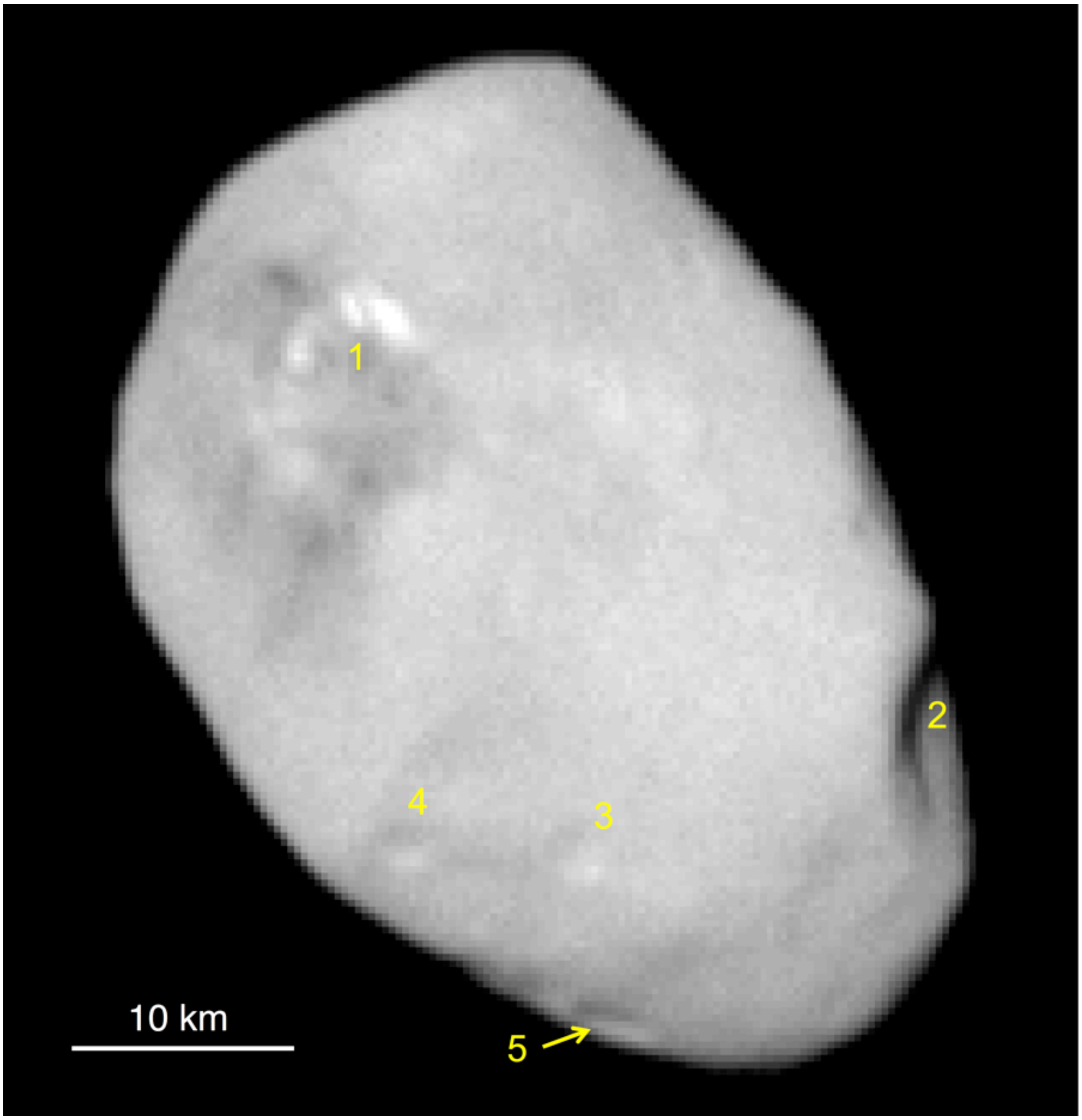

**Fig. S12. Crater locations on Nix.** This is the best resolved image of Nix (raw image taken from observation N_LORRI_LEISA_BEST; see Table 1) with the features we associate with crater impacts marked by yellow ID numbers. The estimated crater sizes and the area used for the crater density calculation are provided in Table S3. Celestial north is up and east is to the left.

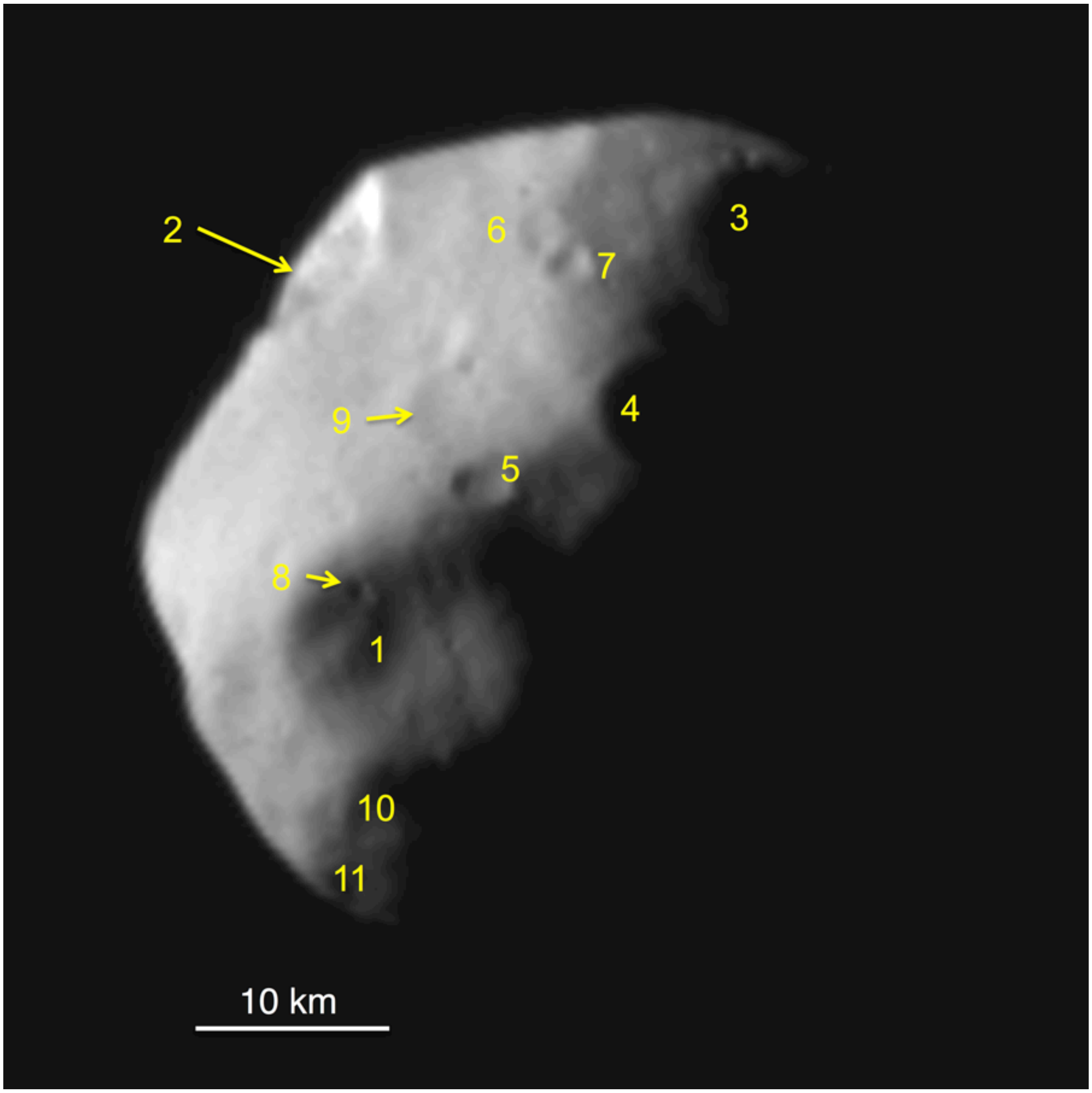

**Fig. S13. Crater locations on Nix.** In this high phase angle (85.9°) image of Nix (2 times super-sampled and unsharp-masked image taken from observation N_MPAN_CA; see Table 1), the features we associate with crater impacts are marked by yellow ID numbers. The estimated crater sizes and the area used for the crater density calculation are provided in Table S3. Celestial north is up and east is to the left.

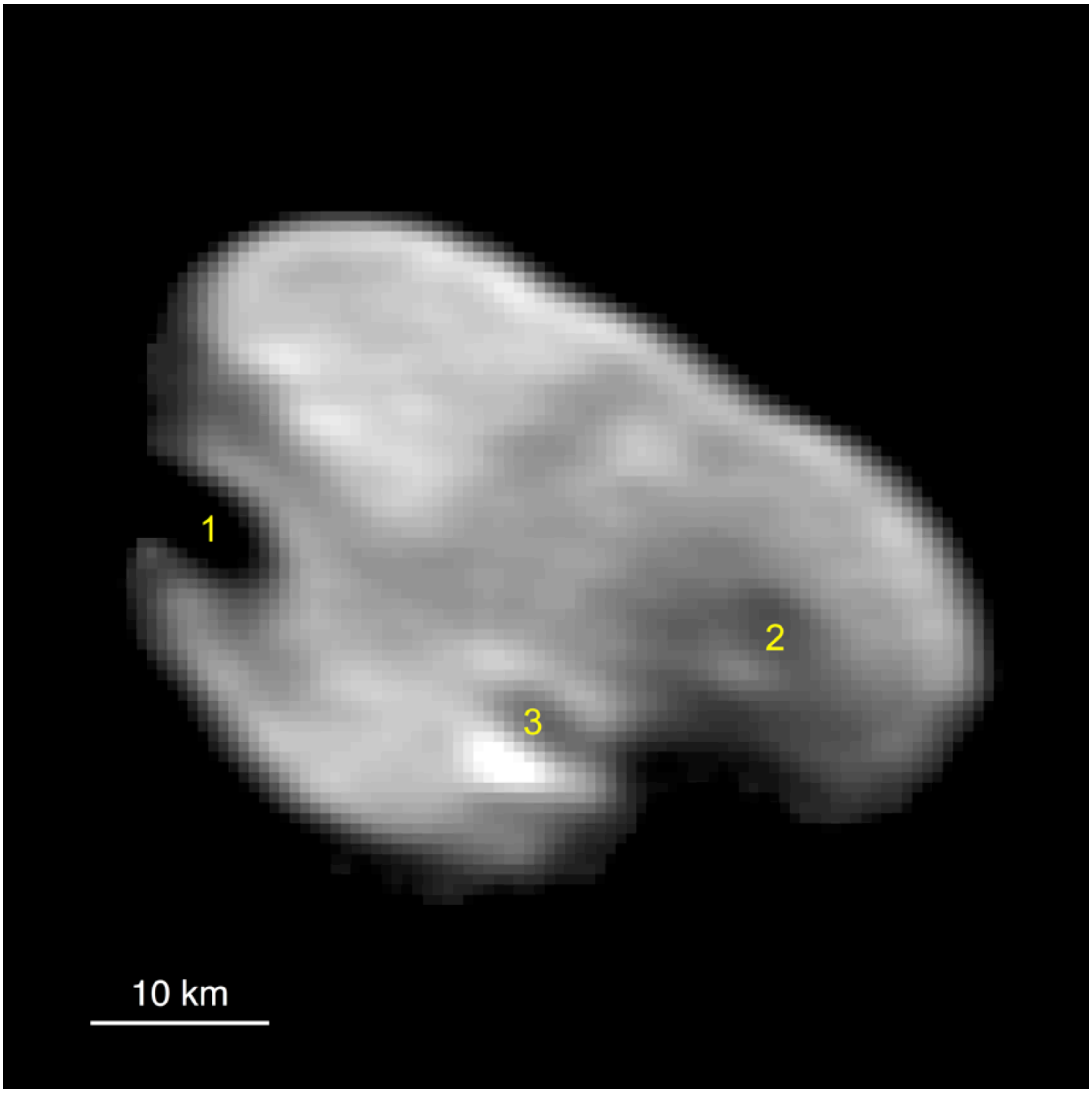

**Fig. S14. Crater locations on Hydra.** This is the best resolved image of Hydra (deconvolved image taken from observation H_LORRI_ BEST; see Table 1) with the features we associate with crater impacts marked by yellow ID numbers. The estimated crater sizes and the area used for the crater density calculation are provided in Table S3. Celestial north is up and east is to the left.